\documentclass{aa}  

\usepackage{graphicx}
\usepackage{txfonts}
\usepackage{amssymb}
\usepackage{amsmath}
\usepackage{graphicx}
\usepackage[colorlinks=true, allcolors=blue]{hyperref}
\usepackage{booktabs}
\usepackage{amsfonts}     % matemaattiset merkit
\usepackage[T1]{fontenc} %font encoding
\usepackage[utf8]{inputenc} %unicode
\usepackage{multicol}
\usepackage{xcolor}
\usepackage{lineno}
\usepackage{gensymb}
\usepackage{soul}
\usepackage{rotating}
\usepackage{comment}
\usepackage{bm}
\newcommand{\eqb}{\begin{eqnarray}}
\newcommand{\eqe}{\end{eqnarray}}

\begin{document}

   \title{Quantitative comparisons of VHE gamma-ray blazar flares with relativistic reconnection models}

   \subtitle{}

   \author{J. Jormanainen
          \inst{1}\fnmsep\inst{2}\fnmsep\thanks{jesojo@utu.fi},
          T. Hovatta\inst{1}\fnmsep\inst{3},
          I. M. Christie,
          E. Lindfors\inst{1},
          M. Petropoulou\inst{4},
          I. Liodakis\inst{1}
          }

   \institute{Finnish Centre for Astronomy with ESO, University of Turku, Finland
         \and
             Department of Physics and Astronomy, University of Turku, FI-20014, Finland
         \and
            Aalto University Mets\"ahovi Radio Observatory, Mets\"ahovintie 114, FI-02540 Kylm\"al\"a, Finland
        \and
            Department of Physics, National \& Kapodistrian University of Athens, University Campus, Zografos, 15784, Greece
             }

   \date{Received; accepted}

% \abstract{}{}{}{}{} 
% 5 {} token are mandatory
 
  \abstract{The origin of extremely fast variability is one of the long-standing questions in the gamma-ray astronomy of blazars. While many models explain the slower, lower energy variability, they cannot easily account for such fast flares reaching hour-to-minute time scales. Magnetic reconnection, a process where magnetic energy is converted to the acceleration of relativistic particles in the reconnection layer, is a candidate solution to this problem. In this work, we employ state-of-the-art particle-in-cell simulations in a statistical comparison with observations of a flaring episode of a well-known blazar, Mrk 421, at very high energy (VHE, E > 100 GeV). We tested the predictions of our model by generating simulated VHE light curves that we compared quantitatively with methods that we have developed for a precise evaluation of theoretical and observed data. With our analysis, we can constrain the parameter space of the model, such as the magnetic field strength of the unreconnected plasma, viewing angle and the reconnection layer orientation in the blazar jet. Our analysis favours parameter spaces with magnetic field strength $0.1$ G, rather large viewing angles ($6$ --$8\degree$), and misaligned layer angles, offering a strong candidate explanation for the Doppler crisis often observed in the jets of high synchrotron peaking blazars.}


   \titlerunning{Comparing VHE blazar flares with reconnection models}
   \authorrunning{Jormanainen et al.}

   \maketitle
%
%-------------------------------------------------------------------

\section{Introduction}
\label{intro}

Blazars are a type of radio-loud active galactic nuclei (AGN) possessing a relativistic jet closely aligned with our line of sight \citep{Blandford1978,Urry1995}. Due to their unique alignment and the relativistic jet speeds, blazars are highly variable sources across the whole electromagnetic spectrum \citep{Scarpa1997,Blandford2019,Hovatta2019}. These variations have been observed to occur in time scales of years all the way down to a few hours or even minutes \citep{Marscher2008,Marscher2010,Jorstad2010,Ahnen2016a,Nilsson2018}. Several mechanisms have been suggested to explain the observed variability of these sources, such as shocks \citep{Marscher1985,Hughes1985,Spada2001,Joshi2007,Graff2008,Liodakis2022, DiGesu2022} and stochastic acceleration by turbulence \citep{Virtanen2005}. While they manage to explain well the slower flares in the lower energies, these mechanisms alone cannot explain the fastest variations detected in the very high energy (VHE, E $>$ 100 GeV) gamma-rays \citep{Aharonian2007,Albert2007} because, typically, the variability resulting from these mechanisms does not reach the intra-night time scales that we observe. For these extreme flares, magnetic reconnection has been suggested as an explanation \citep{Giannios2009a,Giannios2013a}.

In this paper, we consider the magnetic reconnection model presented in
\cite{Christie2019} as the physical mechanism behind this extreme blazar variability. In the reconnection model, initial instabilities of the magnetic field create current sheets within the jet \citep{Spruit2001,Giannios2006,Duran2017,Gill2018} which, in turn, are susceptible to tearing instabilities, thereby allowing the current sheet to fragment into a chain of magnetic islands, or plasmoids \citep{Loureiro2007,Uzdensky2010,Fermo2011,Huang2012,Loureiro2012,Takamoto2013}. These plasmoids, each containing relativistic particles and magnetic fields, are thought to be the origin of blazar variability in the VHE gamma-rays. To realise this scenario, \cite{Christie2019} combined the results of the two-dimensional particle-in-cell (2D PIC) simulations \citep{Sironi2016,Petropoulou2016} with a leptonic radiative transfer code.

In the past, the observed data has been used to estimate the accuracy of the physical models that could account for the blazar variability,  \citep[e.g.][]{Acciari2020, meyer2021} and these kinds of studies can lead us to the right direction. Many properties of blazar jets can already be uncovered through very-long-baseline interferometry (VLBI), which by mapping the inner jet structure, can give estimates of the apparent jet velocity \citep{Jorstad2017,Lister2019,Weaver2022} and estimates of magnetic field strength \citep{2012A&A...545A.113P}. Through single-dish observations it is also possible to constrain the Doppler factor and the bulk Lorentz factor \citep[e.g.][]{Liodakis2018}. Due to their multiwavelength nature, blazars offer also a unique window to the jet via spectral energy distribution (SED) modelling, which allows to estimate the jet power, $\gamma _{max}$, and the magnetic field strength \citep[e.g.][]{2010MNRAS.401.1570T,2014Natur.515..376G}. However, problems arise in such sources where these aforementioned methods are in disagreement \citep[see discussion in][]{Ghisellini2005} and therefore require additional constraints. In addition to elaborating on one possible model to describe the VHE variability, one of the key aims of our study is to overcome these discrepancies by a thorough comparison of first-principles magnetic reconnection simulations and the observed data. In our methods, we engage with the limitations set by the observed data in two steps. First, we use values from literature to set up some of the free simulation parameters to ranges that correspond to our current understanding of these sources. In the next step, we compare the resulting simulated light curves with the observed data in various ways descriptive of our data sets. For the development, we have used only one source, Mrk 421, and present the results of our analysis for one light curve of this source in this paper. In the future, we aim to produce similar studies of different sources in different energies and time scales.

This paper is structured as follows. In Sect. \ref{data}, we describe the observed data set outlined in this introduction. In Sect. \ref{sims}, we give a brief description of the theory behind the simulations and explain the simulation setup in detail. In Sect. \ref{treatment}, we explain the steps taken to treat the simulated data before the comparison. In Sect. \ref{methods}, we give a detailed description of the methods that we developed for this comparison. In Sect. \ref{results}, we state the findings for each subset of simulations. In Sect.\ref{discussion}, we discuss the consequences of these results, while concluding with our findings in Sect.\ref{concl}.

\begin{figure*}
\centering
\includegraphics[scale=0.5]{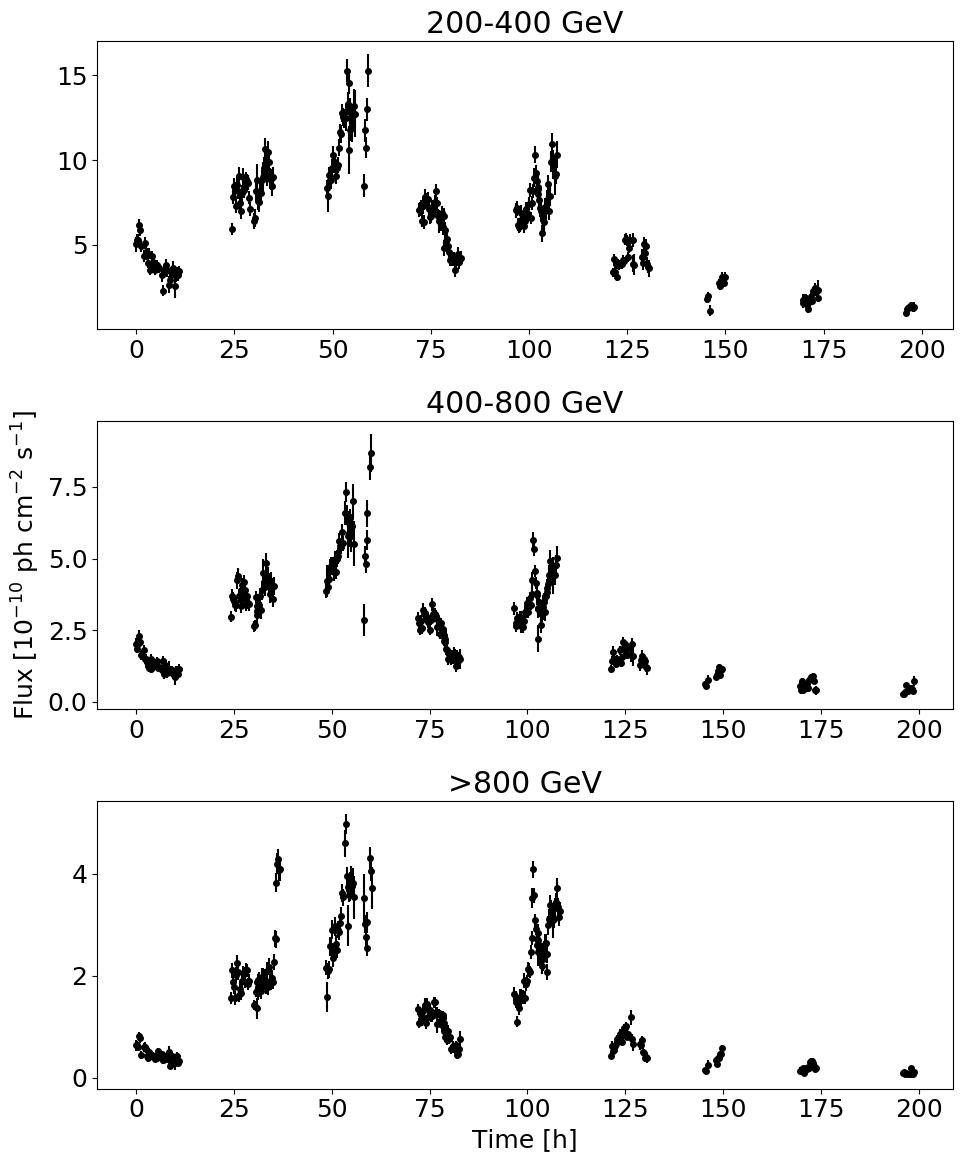}
  \caption{Observed light curves of Mrk 421 in three energy bands of 200-400 GeV, 400-800 GeV, and $>$800 GeV obtained with MAGIC and VERITAS telescopes in 2013 between MJD 56392.8 - 56401.2 \citep{Acciari2020}.}
  \label{fig:observed}
\end{figure*}

\section{Data}
\label{data}

Markarian 421 (Mrk 421) is a bright and nearby BL Lac object at $z = 0.0308$, and thus, a frequently observed source in the VHE domain. The observed data that we use in this paper consists of the VHE gamma-ray light curves of Mrk 421 observed between 11th and 19th of April 2013. These data were observed with two imaging atmospheric Cherenkov telescopes (IACTs) Major Atmospheric Gamma Imaging Cherenkov Telescopes (MAGIC) and Very Energetic Radiation Imaging Telescope Array System (VERITAS) in a simultaneous observing campaign when the source was in an exceptionally active state in both X-rays and gamma-rays \citep{Acciari2020}. The data consists of nine consecutive nights with MAGIC observing in nine nights and VERITAS in six nights, respectively. The strong signal allowed us to divide the data into three separate energy bands of 200-400 GeV, 400-800 GeV, and $>$800 GeV. These data are available online\footnote{https://cdsarc.cds.unistra.fr/ftp/J/ApJS/248/29/fig2.dat} Figure~\ref{fig:observed} shows the light curves obtained via this observing campaign. Because no other VHE blazar has been observed with such a dense temporal cadence across as many nights, many of which have intra-night variability, this unique data set was selected for the development of our method. Furthermore, the source fluxes of Mrk 421 during the quiescent state are known to be around 0.1 Crab units that, for simplicity, can be assumed to be negligible during this flaring event, and we do not assume any other underlying sources of emission for this light curve in our analysis.

Our aim is to compare the predictions of a reconnection model about time scales of the flares, flux amplitudes, and spectral properties with those derived from this unique data set. For our analysis, these data were otherwise kept the same as in \cite{Acciari2020}, but due to some overlapping observations of MAGIC and VERITAS there were duplicate data points that were ignored in our analysis. We also decided to select only strictly simultaneous data points from each band because in the higher energies there exist more data points due to the atmospheric effects affecting the lower energy observations in high zenith angles\footnote{The energy threshold of the observations increases with the increasing zenith angle.}.

\section{Model description}
\label{sims}

Magnetic reconnection has been shown to be an efficient mechanism of accelerating particles to high energies in magnetically dominated plasmas \citep[see][]{Nalewajko2020,werner2021,Haocheng2022,sironi2022} and references therein). In blazars, it has been suggested as a mechanism that could account for the fastest variability observed in the VHE gamma-ray regime \citep[typically 100 GeV - 100 TeV,][]{Giannios2009a, Giannios2013a}. In this study, we consider one such model presented in \cite{Christie2019} in comparison with the observed data. In this section, we summarise the key points of their model and outline the details of the simulation setup utilised in producing the simulated light curves (see Sect. \ref{setup}).

Using 2D PIC simulations, \cite{Sironi2015} showed that magnetic reconnection is able to account for the efficient dissipation of magnetic energy, the extended non-thermal distribution of relativistic particles, as well as the creation of plasmoids which are characterised by a rough equipartition between their magnetic fields and relativistic particles. This work was continued in \cite{Sironi2016}, where the authors reported the statistical properties, namely the distributions of plasmoid size and velocity, of the plasmoid chain in electron-positron pair plasmas\footnote{For results about different plasma compositions, e.g. electron-positron-proton, see \citep{petro2019}}. Additionally, the authors demonstrated that long spatial and temporal scales are required in order to gain sufficient statistical information of the plasmoid chain. These PIC results were incorporated by \cite{Petropoulou2016} into a leptonic radiative model describing the evolution of the radiating particles within a single plasmoid, providing a physically motivated model for plasmoid-powered blazar flares, including the sub-hour and ultra-luminous flares which are characteristic of blazars. Additionally, the authors also derived parametric scalings of the peak luminosity and the flux doubling time-scale as a function of plasmoid size and momentum.

\begin{figure*}
\centering
\includegraphics[scale=0.40]{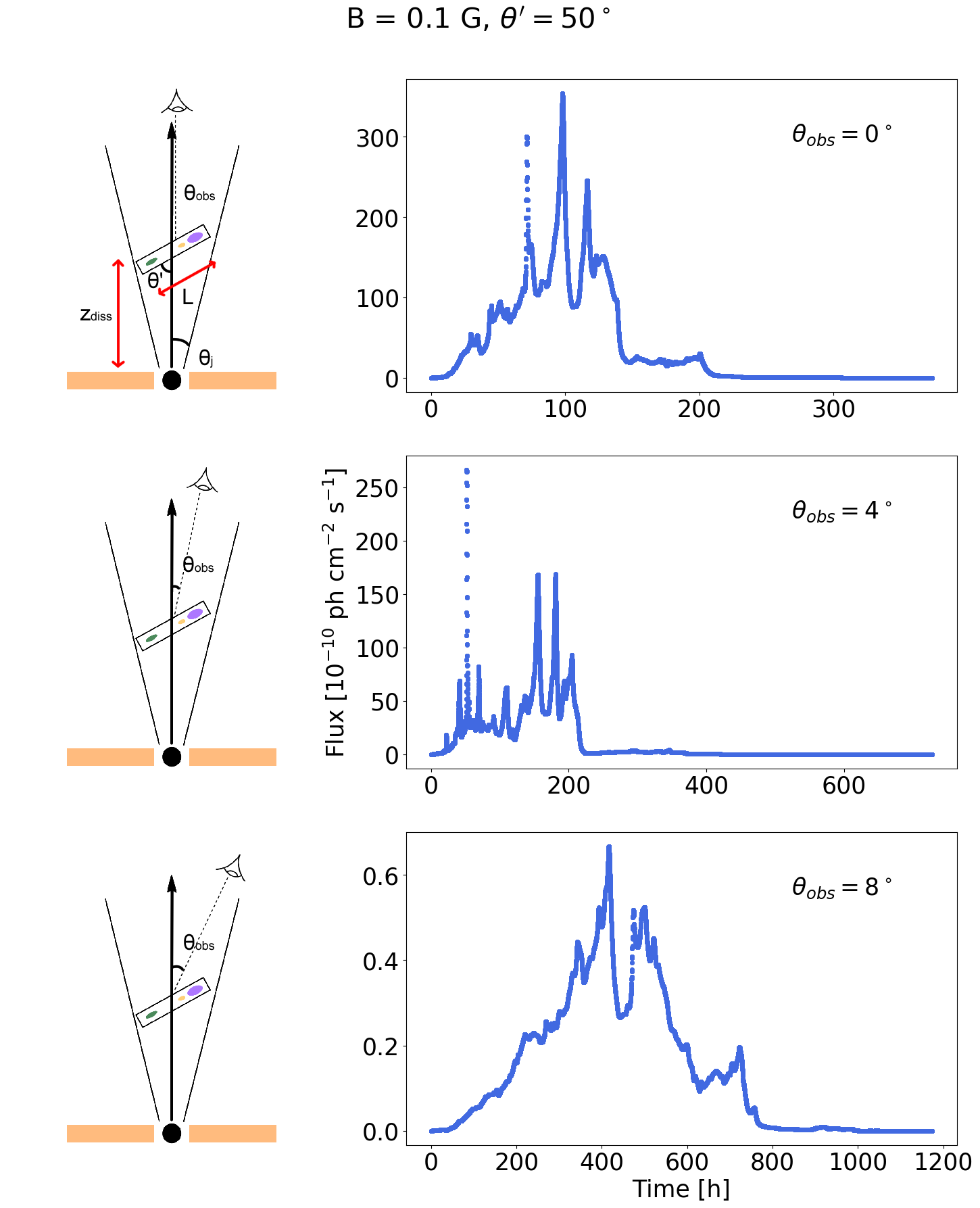}
  \caption{Illustration of the jet schematic, the first panel including the essential parameters of our model, and the resulting light curve when the viewing angle $\theta _{obs}$ is changing, while the reconnection layer angle $\theta '$ and the magnetic field strength $B$ remain constant. Increasing $\theta _{obs}$ results in decreasing fluxes with the Doppler boosting having less of an effect, and increasing the observed time of the reconnection event. Jet schematic adapted from \cite{Christie2019}.}
  \label{fig:jet_thobs}
\end{figure*}

Expanding on the work of these previous studies, \cite{Christie2019} determined the emission of the entire plasmoid chain as applied to the multi-wavelength variability observed in blazars, including both BL Lac and flat-spectrum radio quasar (FSRQ) objects. They developed a time-dependent leptonic radiative transfer model which tracks the evolution of the particle and photon spectra within each plasmoid. This radiative model was then combined with the statistical properties of the plasmoid chain, as derived from the 2D electron-positron PIC simulations of \cite{Sironi2016}, into a simplified blazar model with minimal constraints. As depicted by the first sequence of the schematic in Fig.~\ref{fig:jet_thobs}, the reconnection layer of the adopted PIC simulations is appropriately scaled (see subsequent section for more details) and is taken to be residing within a relativistic jet, with bulk Lorentz factor $\Gamma_j$ and half-opening angle $\theta_j \sim \alpha_j /\Gamma_j$ (where $\alpha_j$ is a scaling factor), at a distance of $z_{diss}$ from the supermassive black hole (SMBH). By varying the observer angle $\theta_{obs}$ and the relative angle between the layer and jet ($\theta^\prime$; i.e. as measured in the jet's co-moving frame), the authors demonstrated that plasmoid chains were able to produce long-duration (i.e. $\gtrsim$ days) light curves which contained the multi-wavelength and multi-timescale variability characteristic of many blazars. 

\begin{figure*}
\centering
\includegraphics[scale=0.40]{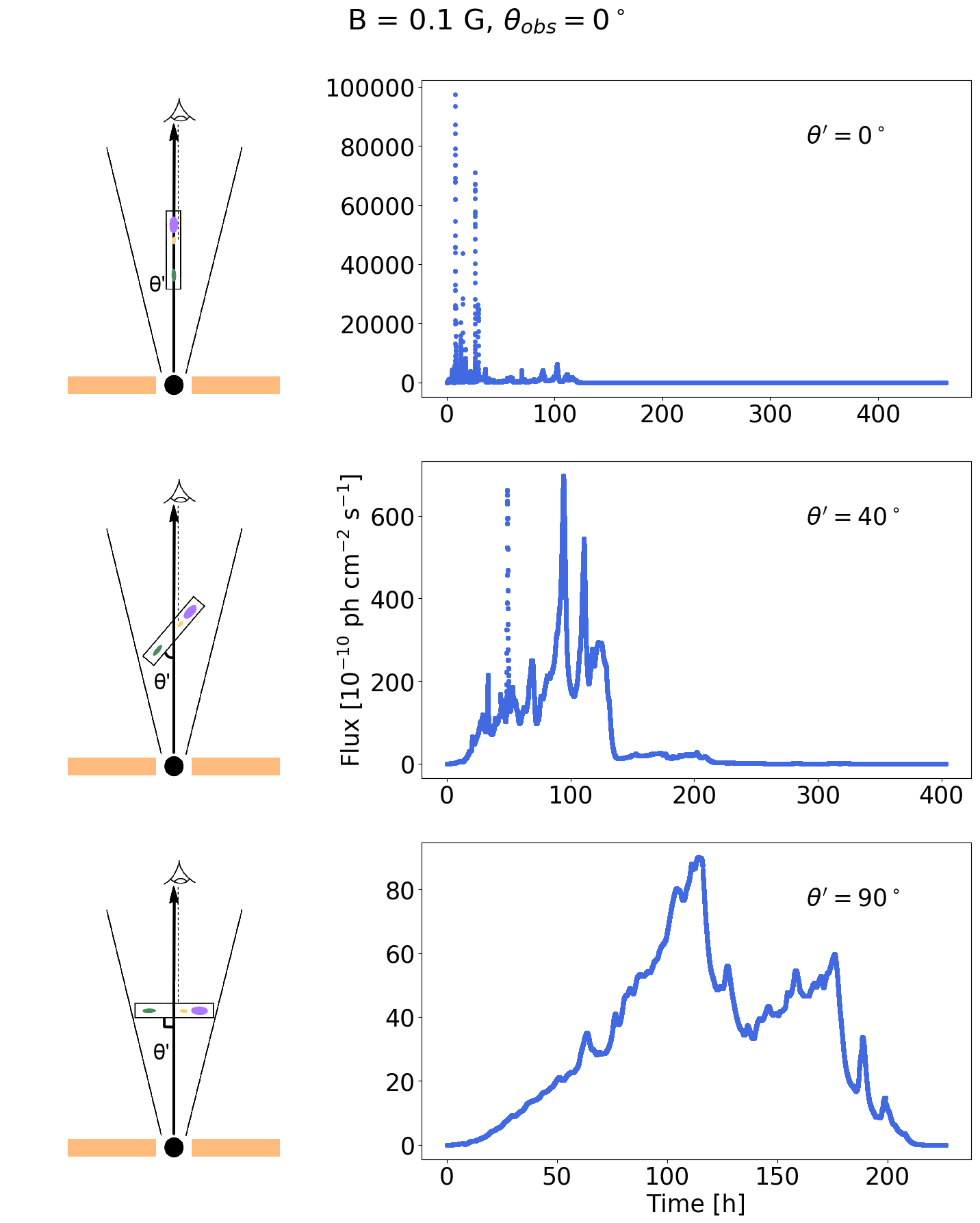}
  \caption{Illustration of the jet schematic and the resulting light curve when the reconnection layer angle $\theta '$ is changing, while the viewing angle $\theta _{obs}$ and the magnetic field strength $B$ remain constant. With the perfect alignment of $\theta '$ and $\theta _{obs}$ we obtain maximum Doppler boosting, resulting in very high fluxes. For increasingly misaligned reconnection layer orientations, the effect of the boosting is diminished. The relative orientation of $\theta '$ has a greater effect on the boosting of the observed emission than $\theta _{obs}$. In addition, with aligned layer orientations the resulting light curve consists of the superpositions of the large and small plasmoids, possessing high amplitude flares with short time scales. With more misaligned layer orientations, the shape of the light curve is dominated by the large-sized plasmoids possessing little short-term variability. Jet schematic adapted from \cite{Christie2019}.}
  \label{fig:jet_thp}
\end{figure*}

\begin{figure*}
\centering
\includegraphics[scale=0.40]{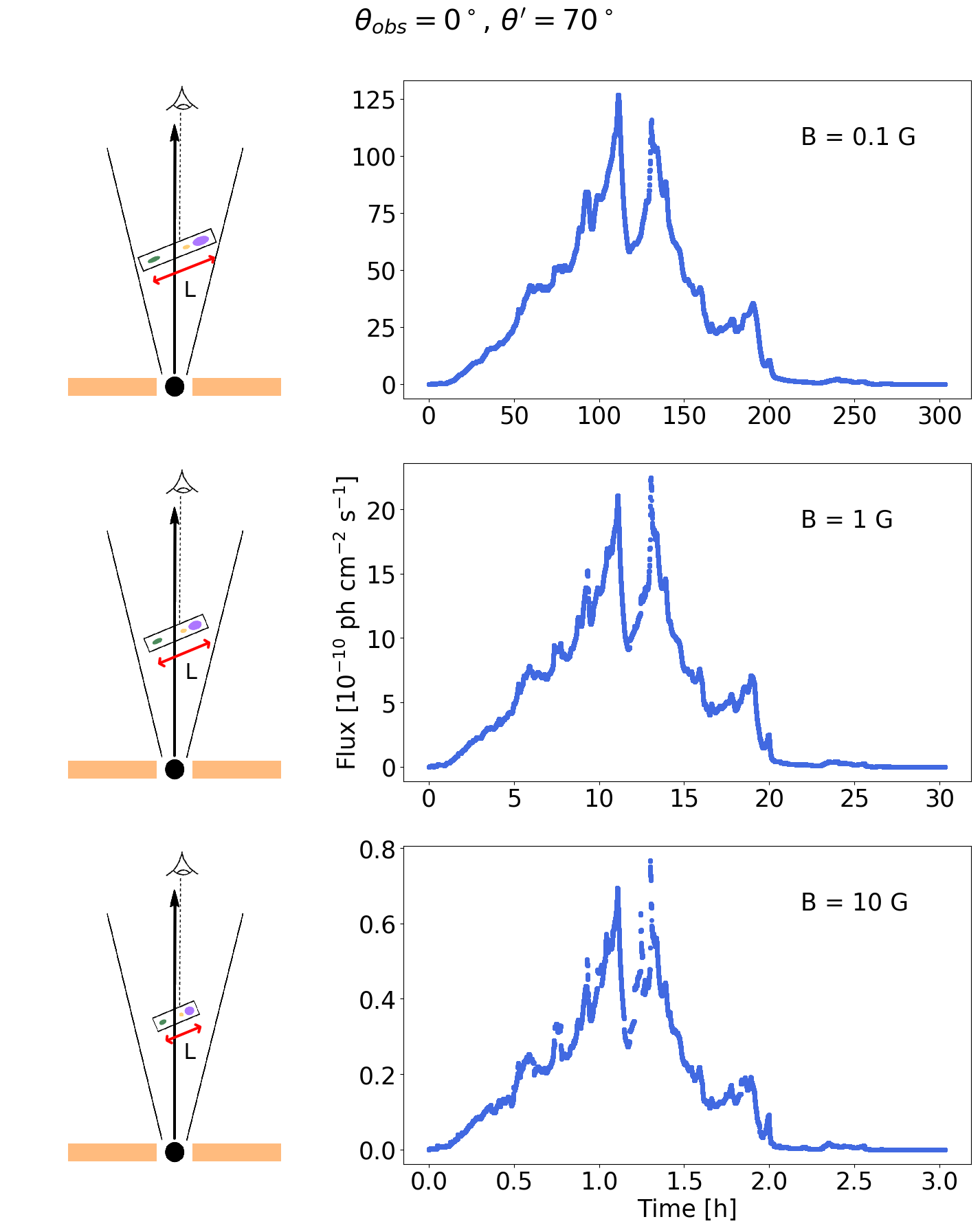}
  \caption{Illustration of the jet schematic and the resulting light curve when the magnetic field strength $B$ is changing, while the viewing angle $\theta _{obs}$ and the reconnection layer angle $\theta '$ remain constant (at $0\degree$ and $70\degree$, respectively). In our model, in order to maintain a constant jet power $P_{j}$, we scale the length of the layer inversely with the increasing $B$. This results in decreasing the observed time of the reconnection event with increasing $B$. Higher $B$ also results in a decreased contribution of the higher energy, inverse-Compton (IC) or synchrotron self-Compton (SSC) component, thus also decreasing the fluxes.  Jet schematic adapted from \cite{Christie2019}.}
  \label{fig:jet_b}
\end{figure*}

\section{Simulation setup}
\label{setup}

To apply the reconnection model of \cite{Christie2019} to a specific source, one needs to adjust several of the model parameters, like the jet Lorentz factor and jet power. Below, we motivate our choices for such model parameters, keeping in mind that we do not aim at fitting a specific data set, but rather find models with comparable fluxes and time scales to the observed ones.

The estimates of the bulk Lorentz factor obtained from SED fitting and VLBI observations differ drastically for Mrk 421. A theoretical upper limit of $\Gamma_j = 4$ has been derived for sources with no detected apparent velocities via VLBI \citep{Piner2018}. In a previous work we used $\Gamma_j = 4$, but this resulted in fluxes that were much lower than those in the observed light curves (see Sect. \ref{distr} for discussion and \citet{Jormanainen2021} for further details). Here, we adopt a slightly higher value, $\Gamma_j = 12$.
    
Another model parameter is the magnetisation\footnote{Defined as $\sigma = B_{up}^2 / 4 \pi m_p n_{up} c^2$, where $B_{up}$ and $n_{up}$ are the magnetic field strength and particle number density far upstream from the reconnection layer, respectively, and $m_p$ is the proton mass. The magnetic field strength within the plasmoid $B_{pl}$ is related to $B_{up}$ as $B_{pl} \approx B_{up} \sqrt{2}$.} $\sigma$ of the jet's plasma at the region where reconnection is triggered. Our model is based on the PIC simulations of \cite{Sironi2016}, therefore we are constrained to the magnetisation values considered therein, that is $3$, $10$, and $50$. For $\sigma \ge 1$ plasmoids accelerate to relativistic speeds (in the jet comoving frame). The asymptotic plasmoid Lorentz factor is $\Gamma_{pl, \max} \sim \sqrt{\sigma}$ for $\sigma \gg 1$ \citep{Sironi2016}. This, when combined with the relativistic bulk motion of the plasma in the jet, can --  under certain orientations -- yield fast flares without requiring very high jet Lorentz factors \citep[see also][]{Petropoulou2016}. For these reasons, we adopt $\sigma=50$ in this work. Moreover, as discussed in \citet{Sironi2016}, $\sigma$ is related to the shape of the injected particle spectrum within each plasmoid, which for $\sigma \gg 10$ can be described as a power-law with slope $p = - d\log N/d\log \gamma \sim -1.5$.

For plasma magnetisation $\sigma \gtrsim 10$, the majority of the particle energy will be stored within the particles with the highest energy \citep[i.e. $\gamma_{max}$; see][]{Sironi2015}. Therefore, we adopt a single minimum Lorentz factor of the injected particle distribution $\gamma_{min} = 500$ which is common to all plasmoids, a value consistent with previous modelling of flares.
    
The peak frequency of the observed synchrotron spectrum is also known to vary, typically moving to higher energies during the flaring states. Since we do not have the estimate for exactly this period, we adopted $\nu_{peak} \sim 1.7 \times10^{16}$~Hz as indicative value \citep{Nilsson2018}. This frequency is used to directly determine the maximum Lorentz factor of the injected particle distribution $\gamma_{max}$, within each plasmoid, since the slope of the injected particle distribution is $p < 2$ (see previous paragraph). This can be estimated as $\gamma_{max} \approx \sqrt{\nu_{peak} / (\delta_{pl} \nu_{syn})}$, where $\delta_{pl}$ is the plasmoid's Doppler factor \citep[as measured by an observer; see eqn.~7 in][]{Christie2019} and $\nu_{syn} = 3 q B_{pl} / (4 \pi m_e c)$, where $B_{pl}$ is the area-averaged magnetic field strength within the plasmoid.

For Mrk 421, the observer angle $\theta_{obs}$ is poorly constrained and was therefore limited to a range of $\theta_{obs} = 0-8\degree$.  In addition, the angle the reconnection layer makes with the jet axis $\theta^\prime$ has not been constrained in the past and was given a range between $\theta^\prime = 0-180\degree$. By performing this variational study of simulating reconnection-driven light curves over a range of values, we can rule out certain orientations of the reconnection layer. The changes to the resulting light curve are demonstrated in Fig. \ref{fig:jet_thobs} for the changes in $\theta _{obs}$ and in Fig. \ref{fig:jet_thp} for the changes in $\theta '$.

As for the final few free parameters, namely the magnetic field strength within each plasmoid $B_{pl}$, the half-length layer of the reconnection region $L$, and the distance at which reconnection is triggered within the jet $z_{diss}$, these are directly related to and can be extracted from the jet power $P_j$. The power of a two-sided jet at a distance $z_{diss}$ from the SMBH can be approximated as \citep{Celotti2008, Dermer2009}
\eqb
P_j \approx 2 \pi c \beta_j (\Gamma_j \theta_j z_{diss})^2 (p^\prime_j+U_j^\prime) \approx 4 \pi c \beta_j (\Gamma_j \theta_j z_{diss})^2 U_j^\prime,
\eqe
where $\theta_j$ is the opening angle of the jet, and $U_j^\prime$ is the co-moving energy density of the jet, which is approximately equal to the jet pressure. For magnetically driven outflows, the latter can be approximated as $U_j^\prime \approx B_{up}^2 / (8 \pi)$ where $B_{up}$ is the magnetic field strength far upstream from the reconnection layer. Additionally, we assume that the opening angle of the jet $\theta_j$ and $\Gamma_j$ are related by $\Gamma_j \theta_j \sim \alpha_j$. Previous studies of blazar flares used $\alpha_j = 1$. However, here we take $\alpha_j = 0.2$ \citep{Clausen-Brown2013,Pushkarev2017,Jorstad2017}. With these assumptions, the jet power can be approximated to 
\eqb
\label{eq:jetpower}
P_j \approx \frac{1}{2} c \beta_j (\alpha_j \, z_{diss} \, B_{up})^2.
\eqe
Therefore, knowing the jet power allows us to place an upper limit on the value of $z_{diss} \cdot B_{up}$. Plasmoids in magnetic reconnection are characterised by rough energy equipartition between magnetic fields and relativistic particles. In this case, the resulting blazar SEDs have typically Compton ratios much lower than unity. In order to obtain a Compton ratio, that is the ratio of the peak IC spectrum to the synchrotron spectrum, of order unity, one has to introduce a multiplicative constant pre-factor within the particle distribution. This pre-factor is dependent upon the magnetic field strength and was empirically found to be 300, 100, and 25, respectively (see Table~\ref{tab:theory_params}). To justify the use of such an alleviating pre-factor, we identify three main sources of uncertainty related to the acquisition methods and the exact jet power of the flaring epoch \citep[for a discussion, see][]{foschini19}. For example, SED fitting of the average spectra of Mrk 421 with the standard one-zone leptonic model infers a jet power of $P_j \sim 1.55 \times 10^{43}$~erg~s$^{-1}$ \citep{Ghisellini2010}, a value that can be assumed to represent a longer-term average of the jet power. We also estimated the jet power from the VLBI observations of MOJAVE \citep{Lister2019} using the formula from \citet{Foschini14} and note that it results in a similar jet power for the source. The estimates from a flaring epoch \citep[e.g.][]{Aleksic2015}  would suggest that during an increased activity the jet power can increase at least by an order of magnitude. We recognise another source of mismatch in the methods of \citet{Ghisellini2010} where the jet powers are calculated only for one-sided jets whereas our calculations take into account a two-sided jet, introducing a factor of two difference for the required final jet power. Because of these uncertainties, we adopt a higher jet power of $P_j \sim 6\times 10^{44}$~erg~s$^{-1}$. Finally, \citet{Christie2020} suggested that the larger plasmoids of the reconnection layer could act as a source of photon background for the smaller plasmoids to upscatter through the IC mechanism, increasing the high-energy flux by a factor of two to four (see also Sect. \ref{disc:modelcaveats}). Combining these sources of uncertainty, we can obtain jet powers closer to the values required by the pre-factors.

With the above analysis, we consider three values of the magnetic field strength within our study, namely $B_{up} = 0.1$, $1$, $10$~G. Using the adopted jet power, the dissipation distances from the SMBH are determined as $z_{diss} \approx 10^{19}$, $10^{18}$, and $10^{17}$~cm, respectively. With $z_{diss}$ determined, we can place an upper limit on the half-length of the reconnection layer $L$ such that $L \lesssim z_{diss} \cdot \tan\theta_j \sim z_{diss} \theta_j \sim z_{diss} \alpha_j / \Gamma_j$. For the three values of $B_{up}$ considered and with $\alpha_j = 0.2$, the reconnection half-length is $L = 1.7\times 10^{17}$, $1.7\times 10^{16}$, and $1.7\times 10^{15}$~cm, respectively (see Table~\ref{tab:theory_params}). The effects of this relation to the theoretical light curve are demonstrated in Fig. \ref{fig:jet_b}.

In summary, we produced 285 light curves scanning three different magnetic field strength $B$ values, each of which had full combination of viewing angle $\theta _{obs}$ values of 0$\degree$, 2$\degree$, 4$\degree$, 6$\degree$ and 8$\degree$ and reconnection layer angle $\theta'$ values from 0$\degree$-180$\degree$ (in steps of 10 degrees). These theoretical light curves were further divided into the energy bands of 200-400 GeV, 400-800 GeV, and $>$800 GeV according to the observed data.

\begin{table}
\centering
\caption{Upper theoretical estimates of the dissipation distance from the SMBH, the half-length of the reconnection layer, and the particle distribution pre-factor for maintaining equipartition for each of the three assumed magnetic field strengths of the reconnection upstream region.}
\label{tab:theory_params}
\begin{tabular}{@{}cccc@{}}
\toprule
 & $B_{up} = 0.1 \ \mathrm{G}$ & $B_{up} = 1 \ \mathrm{G}$ & $B_{up} = 10 \ \mathrm{G}$ \\
\midrule
$\mathrm{log_{10}}(z_{diss})$ [cm] & $19$ & $18$ & $17$\\ 
\midrule
$\mathrm{log_{10}}(L)$ [cm] & $17.23$ & $16.23$ & $15.23$ \\
\midrule
$U_e / U_B$ -pre-factor  & $300$ & $100$ & $25$ \\
\bottomrule
\end{tabular}
\end{table}

\section{Treating the theoretical light curves before comparison}
\label{treatment}

\begin{figure*}
\centering
\includegraphics[scale=0.28]{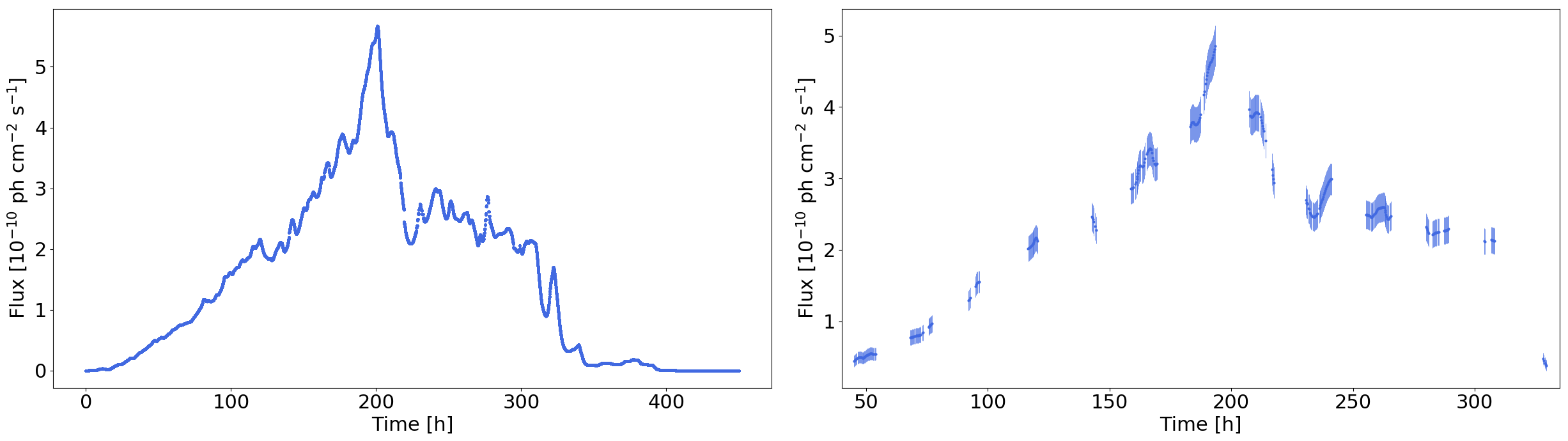}
  \caption{Examples of an untreated theoretical light curve before and after the data has been treated for the comparison. This includes removing the extremely low flux data points, adding flux uncertainties, and matching the observational cadence in terms of temporal resolution and nightly observations.}
  \label{fig:exsims}
\end{figure*}

The theoretical light curves represent a perfect observation of our source without any of the caveats that we face in realistic observational conditions. Because acquiring such a perfect signal is virtually impossible due to the different sources of error and the day-night cycle, the theoretical light curves were treated to resemble the observed data as closely as possible, in other words, the perfect light curves were made imperfect. This includes binning the data, adding error bars, and introducing the daily gaps into the simulated light curves as well as recognising the artificial effects of the simulation process that might otherwise bias our analysis.
Figure~\ref{fig:exsims} shows an example of a theoretical light curve, and the same data after it has been treated to resemble the observations. Below we describe the methods we used to create these "observations" from the simulated data.

\subsection{Removing low flux values}
\label{tails}

An effect of the simulation process that had to be taken into account before the comparison, was the very low values of flux in the beginning and at the end of the theoretical light curves, which we dubbed as "tails". In observed flux units, these values were several orders of magnitude lower than the flux during the reconnection event so a separation between these had to be made.

In the PIC simulations upon which the radiative simulations are based, the evolution of plasmoids and of the contained particles is not traced once plasmoids leave the layer (i.e. simulation box). As a result, in the radiative transfer calculations, the remaining particles within each plasmoid are simply allowed to cool within the ambient magnetic and radiation fields.
Since this might not be an accurate depiction of the flare decaying processes, these decaying parts cannot be accounted for in this model\footnote{For model interpretations of the plasmoid's evolution once it advects from the reconnection layer, see \cite{Petropoulou2016}.}. The same cooling process also introduces the tail features at the end of the theoretical light curves after the flaring episode has died down and the fluxes return to their initial, very low values. These low values also exist at the beginning of the light curve before the reconnection event begins. The shape and the flux of the light curves vary with the orientation of the layer angle in relation to the viewing angle, which determines the Doppler boosting of the plasmoids within the reconnection layer and, thus, largely affects the extension of the tails in contrast with the flaring event. In addition to this, light curves of the different combinations of simulations parameters as well as different energy bands of the same simulation have different flux levels, therefore, a single, universal limit could not be applied to all the simulations. 

\begin{figure*}
\centering
\includegraphics[scale=0.28]{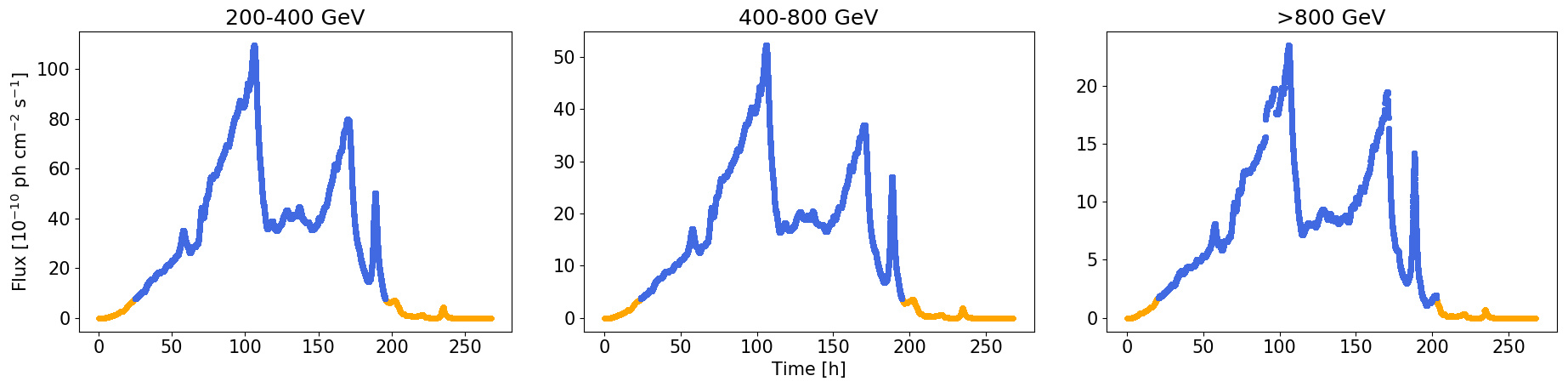}
  \caption{Example plots of a simulation where the tails (in orange) have been identified from the flaring event (in blue) in the beginning and in the end of the light curve in each energy band. The tails are not cut identically in each energy due to differences in the flux levels derived differently for each light curve.% we derive the limit for each light curve separately. 
  This effect is later nullified by only selecting the simultaneous data points for the analysis.}
  \label{fig:tails}
\end{figure*}

In order to differentiate between the flux values of the tail and the reconnection process, the simulated fluxes were normalised and plotted in an optimised histogram using the Scott's rule \citep{SCOTT1979}. A crude cut was made by selecting enough of bins (first four) from the beginning of the histogram so that almost all simulations had their tails cut from the beginning and the end of the light curve in a similar fashion in all three energy bands.
Figure~\ref{fig:tails} shows an example of a light curve where the tails have been identified in contrast with the actual reconnection event. This way, we were able to minimise the number of possible biasing low flux values from these light curves. This was deemed to be the least intrusive way of cutting the tail from each simulation.

As explained in Sect. \ref{data}, we selected only the simultaneous data points of the observed data for these analyses. Because we treat the observed and the simulated data the same way, we selected only the simultaneous data points from the simulated light curves as well. This also nullifies the possible bias of the tail cuts being slightly different in each energy.

\subsection{Binning and sampling}
\label{sample}

In order to make the simulated, ideal light curves resemble our observed data, the theoretical light curves had to be binned into a similar temporal cadence as the observed data and sampled in a way that mimics realistic observations, thus, taking into account the daily gaps.

As explained in Sect. \ref{treatment}, the theoretical light curves do not suffer from the observational conditions that we face in reality, and specifically, in Earth-based observations. In the VHE observations, the data are also averaged over time on such time scales where no significant variability is detected and where the acquired signal exceeds a certain significance threshold. In the case of the light curves of Mrk 421 that we use in this analysis, the data are binned into 15-minute temporal intervals. In addition to this, the observed data always includes some gaps due to the daytime when observations cannot be made, varying weather conditions, and possible technical issues. Therefore, when looking at an observed light curve we cannot be sure of how the source behaves at all times, and if we have observed a full flare when recognising a flare-like structure in the data. 

To account for these limitations, the simulated data first need to be binned into the temporal cadence of 15 minutes and then sampled in a way that only certain parts of the theoretical light curve are observed at a time, in a similar manner as the observed light curve that has gaps due to daytime. We introduced daily gaps into the theoretical light curves by using the exact observed cadence in the selection of the data points from the theoretical light curves. In order to display different parts of the light curve in each sample, we rotated the daily gaps across the theoretical light curve by shifting the observed times by a randomly added value between 0.5 and 100 hours. In the case of the longest simulations ($>$100 hours), we repeated this procedure a thousand times to obtain a large enough variety of light curves. Therefore, we end up with 1000 realisations ("simulated" light curves from here on) of one theoretical light curve to be compared with the observed data. In some cases, the theoretical light curves were also much longer in duration compared to the observed data that spans
about 200 hours so we needed to cut the light curves that were longer than 300 hours close to the observed 200 hours to make the comparison of these light curves more simple. The shortest simulations ($<40$ hours, therefore shorter than two days) were only either compared as a single night of observations or sampled using a sliding window to a duration of a single night (see Sects. \ref{b1sims} and \ref{b10sims} for details).

\subsection{Deriving the uncertainties}
\label{errors}

Because the simulation process does not introduce similar sources of uncertainty in the produced values of flux as the observed data has, the theoretical light curves are also in this sense ideal observations of the source. Therefore, we needed to include some error estimation of the theoretical fluxes in our analysis in order to make them resemble the observed data. We derived the uncertainties of the theoretical fluxes using the observed uncertainties to avoid further assumptions on the sources and shapes of error. To match the scaling of the uncertainties in relation to the photon flux in the observed data, we calculated a scaling factor that we applied to the theoretical fluxes to obtain similarly behaving uncertainties. We did this by fitting the observed relative error, or flux/error ratio, $f(x)$ against the observed flux $x$ with the function

\begin{align}
\label{shape}
f(x) = A \cdot x^{n}.
\end{align}

\begin{figure}
\centering
\includegraphics[scale=0.55]{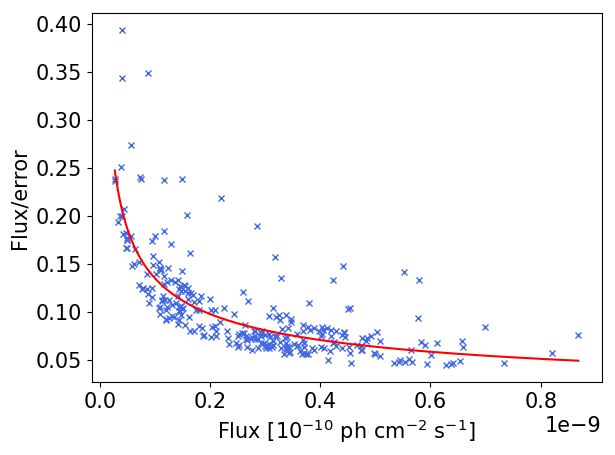}
  \caption{The flux/error ratio against the observed flux for the observed data in the 400-800 GeV band. The red curve shows the fit according to eqn.~\ref{shape}.}
  \label{fig:errors}
\end{figure}

We multiply the simulated fluxes with the flux/error ratio and obtain estimations for the uncertainties. Figure \ref{fig:errors} shows the relation of the flux/error ratio against the observed flux for the observed data in the 400-800 GeV band and the fitted function. The calculated values of $A$ and $n$ for each band are $A_{200-400 \ GeV} = 2.91 \cdot 10^{-6}$, $n_{200-400 \ GeV} = -0.48$; $A_{400-800 \ GeV} = 3.07 \cdot 10^{-6}$, $n_{400-800 \ GeV} = -0.46$; and $A_{>800 \ GeV} = 6.43 \cdot 10^{-6}$, $n_{>800 \ GeV} = -0.52$.

In turn, we did not apply any additional noise on top of the theoretical fluxes since we could not account for the shape and sources of the noise on top of the observed data, that is we avoid biasing the shape of the simulated light curves with excess assumptions.

\section{Methods}
\label{methods}

In this section, the methodology developed for the comparison of the simulated data against the observations is described in detail with examples. The two different perspectives from which we approach our data are the comparison of time scales and the comparison of flux amplitudes, both with various tests. Additionally, for this unique data set where data could be divided into three different energy bands we also compared the spectral properties of these data. All of these tests were used to narrow down the possible parameter space of the model. As we will later show in Sect. \ref{results}, the best matching simulations, the \textit{gold sample} that we will define later in Sect. \ref{b01sims}, are indicated by a combination of almost all of these tests.

This section describes the basic principles behind our analyses that were designed for data sets with a large enough number of points to allow a statistical analysis chain.
In our case, the longest simulations were those with the magnetic field strength $B = 0.1$ G. We give more details about the steps taken to analyse the shorter simulations with $B = 1$ and $10$ G in Sects. \ref{b1sims} and \ref{b10sims}, respectively.

\subsection{Time scales}
\label{times}

As the fast variability of the flux is the most striking observational feature in blazar light curves, it is natural to start the comparison between simulations and data from the time scales. Typically, the time scale analysis is limited to looking for the fastest flux-doubling time in the observed data and possibly searching for a parameter set with a similar flux-doubling time scale. In this work, however, we first tried to exclude simulations that show too fast variability that would have been observable by the current generation IACTs if present in the observations (see \ref{intrabin}). Then we performed a systematic comparison of the rate of variation (i.e. rate of change) within the simulations and compared it to the full observed light curves (see \ref{change}).

\subsubsection{Intrabin variability}
\label{intrabin}

As explained in Sect. \ref{sample}, the theoretical light curves were binned into similar temporal intervals as the observed data, and most of the tests that we perform on these data are done for the data that has been binned accordingly. However, before binning the theoretical light curves, we investigated them by eye and noticed that the variability in many of our simulations appeared to be more extreme than the variability observed for Mrk 421. The fastest time scales observed for Mrk 421 so far were estimated in \cite{Acciari2020} where they determined the flux-doubling time scales of each night of this data set. The fastest flux-doubling time scale that they found was $0.098\pm 0.029$ hours in the 400-800 GeV band for the night of 15th of April 2013. The observed data is binned into 15-minute bins, therefore the variability detected for this night is faster than the temporal cadence of the light curve. However, because in the observed light curves of Mrk 421 variability faster than 15 minutes was not detected in all of the bands, and the variability in the other nights is close to the binning of the 15 minutes or slower, we sought for variability that is more extreme within the 15-minute bins of the simulated light curves.

As a first evaluation of the simulated time scales, we looked for the fastest time scales present in the unbinned theoretical light curves. In order to estimate whether such fast variability time scales and amplitudes would be detectable with the current generation of IACTs, we first needed to assess the sensitivity to detect 5$\sigma$ variability in one-minute time scale. The detectability of the fast variations depends on the photon flux and the amplitude of these variations. We calculated the flux limits for the one-minute time scale detections of a source with Crab-like spectrum. These limits were $8.0 \cdot 10^{-11} \ \mathrm{ph/cm^2/s}$ for the 200 - 400 GeV band, $5.4 \cdot 10^{-11} \ \mathrm{ph/cm^2/s}$ for the 400 - 800 GeV band, and $4.1 \cdot 10^{-11} \ \mathrm{ph/cm^2/s}$ for the $>$800 GeV band. Since the spectral shape we used for deriving these limits does not match the SED of our source, these limits are simply indicative limits for the observing capability of MAGIC.  Therefore, for fluxes close to or higher than the derived limits, such fast variability would be detected with MAGIC and other current IACTs. Only for the lowest fluxes of these simulations, this could not be reliably assessed.

Instead of the flux being doubled within the 15-minute bins we set as a criterion to look for those data points where the flux is tripled within the 15-minute bins. Because in some cases the theoretical fluxes are lower than our derived sensitivity limit, any data points that are below the limit are not taken into account in this test. In summary, from the unbinned theoretical light curves, we search for bright large amplitude flux variations that would not go unnoticed in the observed data, and flag the coinciding data points in the binned simulated light curves.

\begin{figure}
\centering
\includegraphics[scale=0.55]{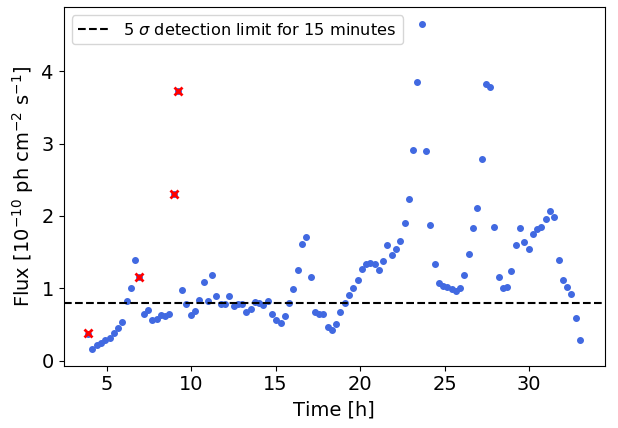}
  \caption{An example of a binned simulated light curve where the data points with fast variability are flagged (red crosses). The black dashed line shows the 5 $\sigma$ detection limit for the given energy range with 15-minute integration time for detecting such variability with the current generation IACTs. Data points below the limit are neglected.}
  \label{fig:intrabin}
\end{figure}

Because in reality we cannot get continuous observations of our source due to daytime, some of this fast variability could go undetected, especially if present in only a small portion of a theoretical light curve. Therefore, we determined the chance of detection of even one case of such extreme variability by calculating the probability of detection from the 1000 samples that utilised the temporal cadence of the observed light curves. We found that with such a cadence this kind of variability would be detected with a chance of 30\%. The high probability of detection is due to untypically good coverage of the observations used here, with a continuous duration of 6-10 hours per night\footnote{The presence of intravariable bins alone is in general not enough to rule out a simulation since the typical VHE gamma-ray observations are not as densely sampled as in our example data set.}. Figure~\ref{fig:intrabin} shows an example of a simulated light curve where the bins with variability faster than 15 minutes are detected and marked as red crosses. For easier identification of the best simulations, we give the inverse of the calculated probability as a final result, highlighting those simulations where the detection of intrabin variability is least likely.

\subsubsection{Rate of change}
\label{change}

A common problem in the fitting of the blazar flares is how to define the flares in these light curves, especially if these flares have not been observed completely, they have been observed with an insufficient cadence, or they overlap with each other. In the VHE gamma-rays, the observed blazar fluxes do not show a flux baseline even during the quiescent state, which also further complicates the estimation of the true amplitude of these flares. Because of these ambiguities, we sought to assess the variability time scales with the least restrictive model that would not assume a flux floor or a defined shape for the flare. In addition, most VHE light curves are rather poorly sampled and, therefore, fitting a model with multiple free parameters would not be feasible either from the perspective of not having enough data points or to actually tell whether we have observed a complete flare. The Bayesian blocks method \citep{Scargle1998, Scargle2013} aims to recognise statistically significant changes in flux without making an assumption about the shape of the flare. In this case, "flares" are simply rising and falling shapes identified by the blocks code, grouped together by the HOP algorithm \citep{Eisenstein1998}, and no further definition is used for them. This was deemed as an as objective as possible way of deriving the rate at which the flux increases in each identified shape. As explained in Sect. \ref{tails}, the decays of the flares are only described by the cooling of the electrons without an additional physical model of the plasmoid evolution upon ejection from the layer. Due to this, we only focus on the rise times of the flares, which are defined by the acceleration of the electrons by the bulk acceleration of plasmoids in the reconnection layer.

\begin{figure*}
\centering
\begin{tabular}{ll}
\includegraphics[scale=0.45]{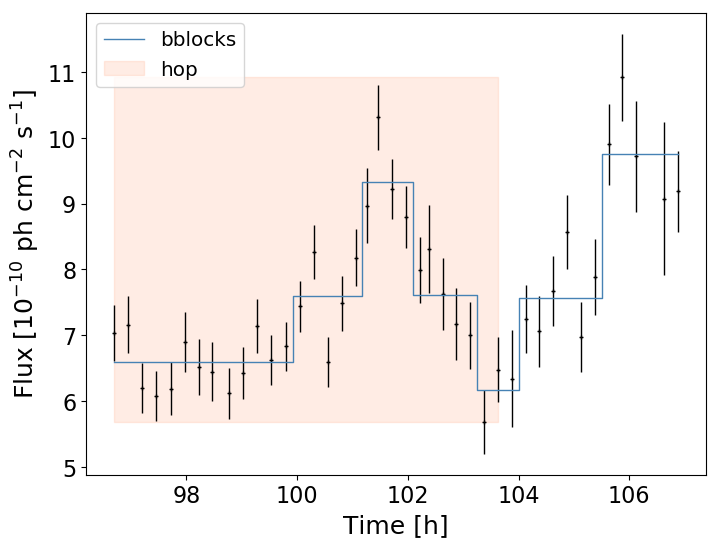}
\includegraphics[scale=0.45]{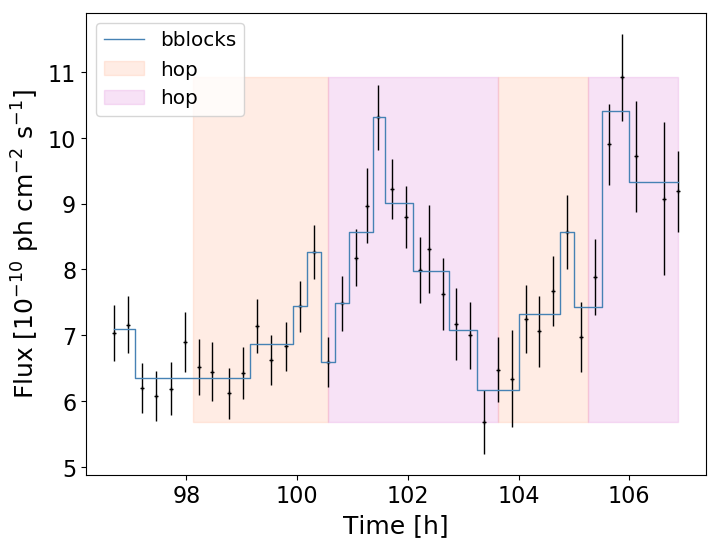}
\end{tabular}
  \caption{Night of April 15 2013 of Mrk 421 fitted with the Bayesian blocks, highlighting the identified flare structures. On the left panel, the $gamma$ value used is 0.1 as used in our analysis. On the right panel, the same light curve fitted with $gamma = 0.5$ showing more flares.}
  \label{fig:blocks}
\end{figure*}

For the fitting, we use a Bayesian blocks code adapted by \cite{Wagner2021} to identify the flares. For the identification, we chose the method "half" where the valley blocks between each flare are simply divided in half. We therefore define the flare rise time as the time from the middle of the valley block to the middle of the highest amplitude block. The flare amplitude is then divided by the rise time, giving us the rate of change of the flare. Because the Blocks code has one tunable parameter $gamma$ that determines the sensitivity of the detected variations, we adjusted this parameter specifically for these data by looking for a $gamma$ value where the identified flares are roughly the same in all three energies and the fitted structures are not clearly overfitted. We used a $gamma$ value of 0.1. Figure \ref{fig:blocks} shows an example of a fitted observed night of Mrk 421 (April 15 2013) with two different $gamma$s, 0.1 and 0.5.
 
\begin{figure*}
\centering
\includegraphics[scale=0.65]{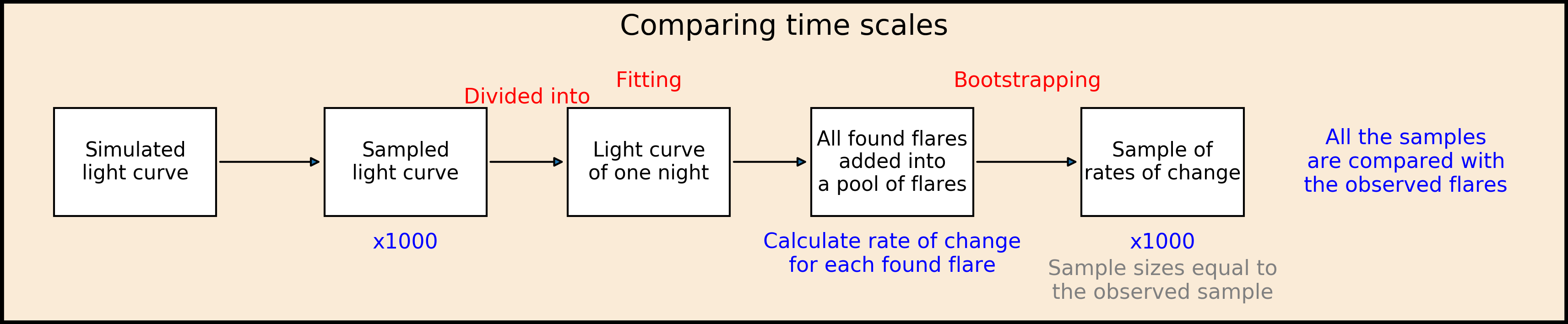}
  \caption{Schematic of the methodology of the time scale analysis.}
  \label{fig:schem}
\end{figure*}

\begin{figure}
\centering
\includegraphics[scale=0.55]{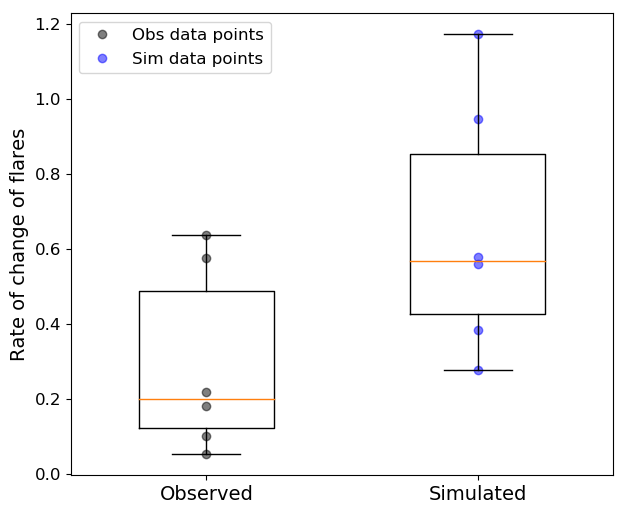}
  \caption{An example of a boxplot showing the range of the rate of change values for the observed and the simulated data. The box represents the IQR, the orange line inside the box is the median, and the whiskers show the spread of the data. Overplotted are also the actual values in black and blue.}
  \label{fig:boxplot}
\end{figure}

Figure~\ref{fig:schem} shows the schematic of the analysis process of the time scales. We divided each sampled light curve nightly to avoid fitting over the daily gaps. From the fits we collected a "pool" of flares representing the possible observed flares of each simulated light curve, and used bootstrapping to draw samples with an equal number of flares than the observed light curves. We compared each bootstrapped sample with the observed sample. Because the number of the found flares from the observed data per light curve was less than 20, using a statistical test that compares these data as distributions was not feasible. Therefore, we searched for the overlap of the interquartile range (IQR) instead. In descriptive statistics, IQR describes the spread of the data as the range between the 25th percentile and the 75th percentile of the data. This means that the data is divided into the upper and lower halves by the median and the 25th percentile is the median of the lower half of the data with the 75th percentile being the median of the upper half respectively. Figure \ref{fig:boxplot} shows an example of a comparison of the observed flares against a bootstrapped simulated sample. While this method still allowed those samples where the spread of the data is large or that have drastic outliers to be matched with the observed sample, these cases were still deemed to be rare enough to not have a drastic effect on our results.

\subsection{Amplitudes}
\label{amp}

Directly following from the flaring nature of the blazar light curves, the next step in the analysis was to compare the amplitudes of the flux variations. We approached this by first comparing the photon flux distributions (see Sect. \ref{distr}) that in addition to the shape of the light curve take into account the general level of the flux in these light curves. Next, we compared the fractional variability factor of the observed light curve with a distribution of fractional variability factors derived from the simulated light curves (see Sect. \ref{fvar}).

\subsubsection{Photon flux distribution}
\label{distr}

One way to compare the simulated flux amplitudes is to look at the distribution of the flux throughout the light curves. A similar study was made in \cite{Jormanainen2021} where we compared the normalised fluxes instead of the raw fluxes with the observed data, but we have updated our method since and will describe the steps taken again in detail.

\begin{figure}
\centering
\includegraphics[scale=0.7]{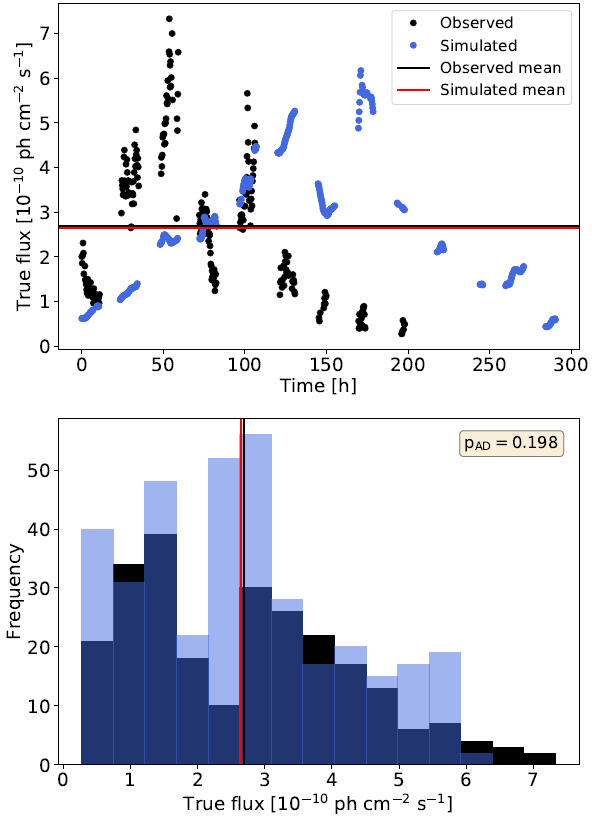}
  \caption{An example of a comparison of the flux distributions. The top panel shows the comparison of the simulated (blue) and the observed light (black) curve together with their means. The bottom panel shows the flux distributions as histograms together with the respective means and the p-value of the AD-test.}
  \label{fig:distr}
\end{figure}

In order to find the simulations that might produce a similar flux distribution as we have in the observed data, we compare the distributions of all 1000 samples of each simulation with the observed light curve separately for each energy band. To estimate the similarity, we use the two-sided Anderson-Darling test and select matches based on the p-values that are larger than 0.05, indicating that with $95 \%$ reliability we cannot reject the null hypothesis that these data sets share the same underlying distribution. Figure~\ref{fig:distr} shows an example of one such comparison. In the top panel, we have visualised the simulated light curve (in blue) and the observed light curve (in black) against each other, and in the bottom panel, we show the flux distributions as well as the p-values of the two statistical tests.

In \cite{Jormanainen2021}, we concluded that the simulations we produced with the bulk Lorentz factor $\Gamma _{j} = 4$ \citep{Piner2018} that was based on the literature value and a smaller half-length of the layer $L$ the resulting fluxes were often 100-1000 times lower than those of the observed data. Because of this, we reran our simulations for this study with an increased $\Gamma _{j}$ and $L$ to obtain fluxes closer to the observed values. Thus, with the current set of the simulations we were able to use the raw simulated fluxes as one criterion when looking for matching light curves based on their flux distributions. In an ideal case, each of the three energy bands would give us a match when sampled in a similar manner, which in turn would indicate that the simulated SED matches the observed data well.

\subsubsection{Fractional variability}
\label{fvar}

We took another approach to assess the flux amplitudes by quantifying and comparing the variability of the simulated and the observed light curves. We did this by comparing the fractional variability factors of the sampled simulations against the observed variability.

The fractional root mean square (rms) variability amplitude, or the fractional variability factor describes the degree of variability of a light curve. In \cite{Edelson2002}, the fractional variability factor is defined as

\begin{align}
F_{var} = \sqrt{\dfrac{S^{2} - \langle \sigma ^{2}_{err}\rangle}{\langle x \rangle ^{2}}},
\end{align}
where $S^{2}$ is the variance of the data set and $\langle \sigma ^{2}_{err}\rangle$ is its mean square error. The error for the fractional variability is given in \cite{Poutanen2008} as
\begin{align}
\Delta F_{var} = \sqrt{F^{2}_{var} + err(\sigma ^{2}_{NXS})} - F_{var}
\end{align}
where the $err(\sigma ^{2}_{NXS})$ is the error of the normalised excess variance as defined in \cite{Vaughan2003}. We computed the fractional variability factors using the $\mathit{compute\_fvar}$ function in the \textit{gammapy}\footnote{https://gammapy.org/} python package.

\begin{figure}
\centering
\includegraphics[scale=0.40]{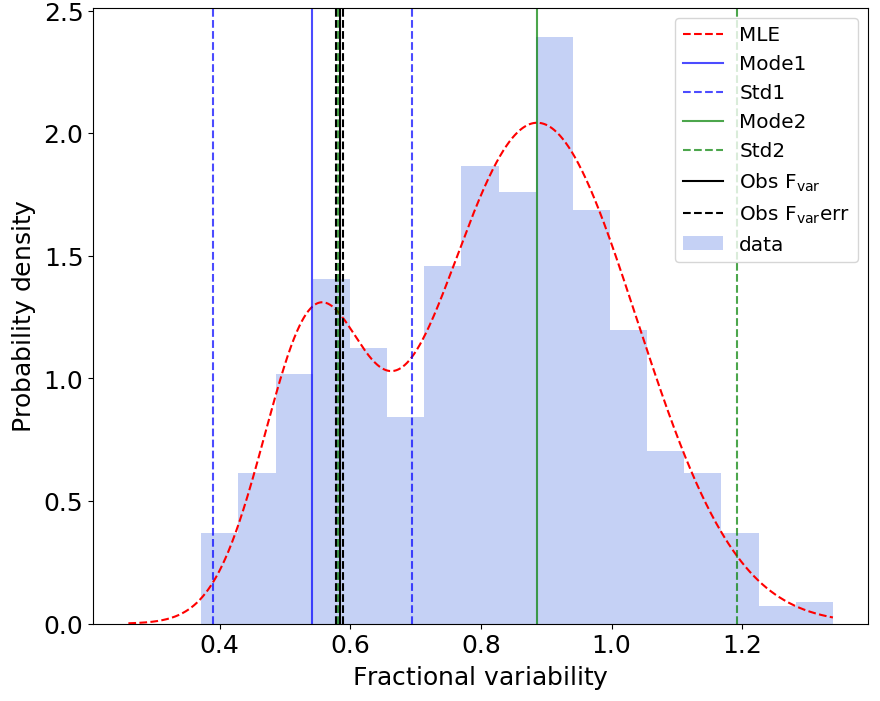}
  \caption{An example plot of the fractional variability comparison where the histogram shows the distribution of the fractional variability factors from all the samples of one simulation. In this case, we fitted the distribution with two Gaussian shapes, and the red dashed line shows the sum of this fit. The black line shows the observed value (errors with black dashed lines) in comparison with the distribution modes and their 2$\sigma$ limits (black and green lines, 2$\sigma$ limits with dashed lines).}
  \label{fig:fvar}
\end{figure}

We compute the fractional factor for each sample of each simulation and combine them into a distribution shown as a histogram in Fig.~\ref{fig:fvar}. We fit this resulting distribution with either a unimodal or a bimodal Gaussian shape in order to get a rough idea of the variability across the different samples of each simulation. The decision between the two fits is made by selecting the fit that gets a lower score of the Bayesian information criterion that elaborates the goodness of fit based on log-likelihood \citep{Liodakis2017,Liodakis2019}. We calculate the fractional variability factor for the observed data and compare it against the simulated distribution. From the Gaussian fit, we derive the mode for the distribution and if the observed value falls within two standard deviations from the mode/modes of a distribution, we accept this simulation as a potential candidate for the underlying parameter space. Again, ideally, all energy bands would give a match if the simulated model describes the observed SED well.

\subsection{Spectral properties}
\label{spectra}

One of the reasons this data set of Mrk 421 was chosen for our analysis was the largely simultaneous flaring data on three different energy ranges that we could use to explore the spectral properties of the simulations like with no other source. The shape of the simulated SED and its accuracy compared to the observed SED has a direct impact on the results of the other tests, especially those concentrating on the flux amplitudes. To compare the SEDs, we computed the slope of the spectrum of each simultaneous data point in the three energy bands. This way, we could estimate the time evolution of the spectra of the light curves, and by combining the obtained values of spectral slopes of each light curve into distributions we were able to roughly compare the observed and the simulated SEDs.

Because we only use the observed light curves in our analysis, the detailed comparison of the simulated and the observed SEDs was beyond the scope of our analysis. Therefore, we did not fit the observed or the simulated spectra with a defined shape, typically a power law or a log-parabola, but calculated a simple slope from the flux data by fitting the flux and the frequency of each band in log-log space with 

\begin{align}
\label{line}
f(x) = m \cdot x + B.
\end{align}

\begin{figure}
\centering
\includegraphics[scale=0.35]{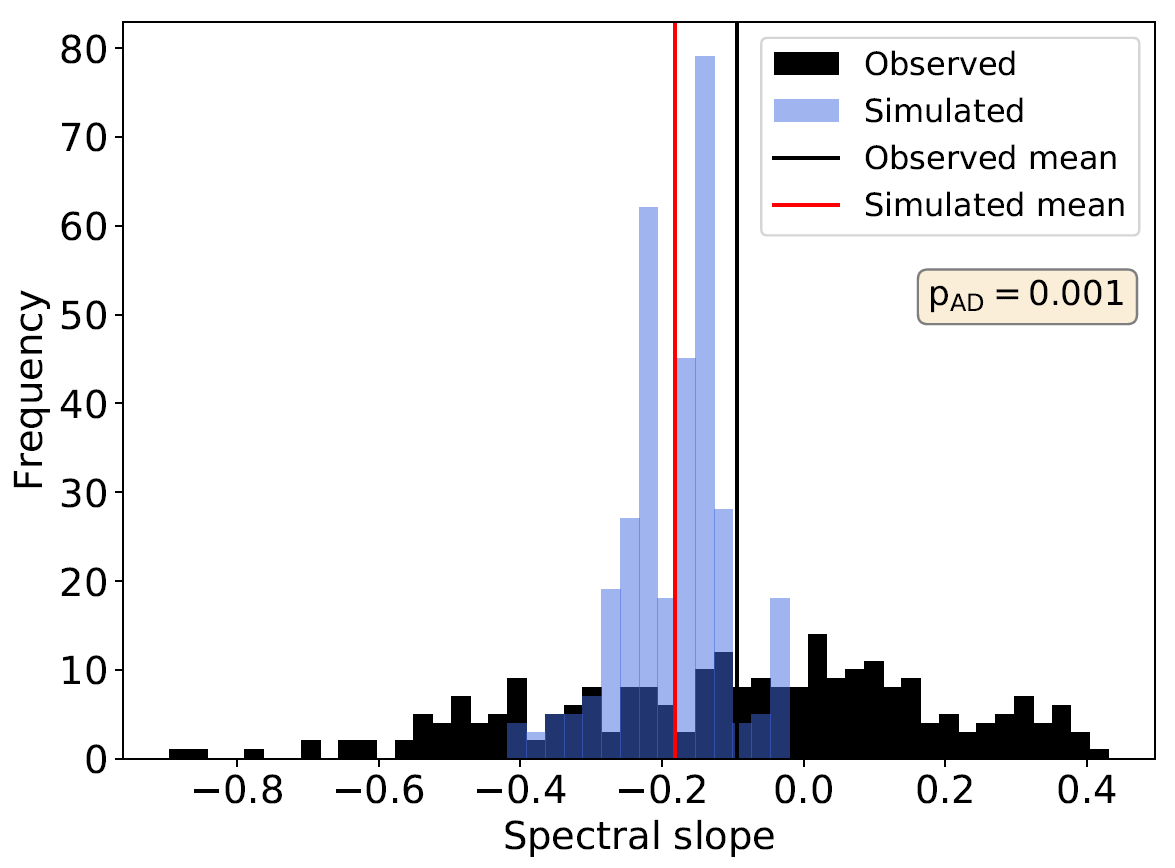}
  \caption{An example of a spectral slope distribution comparison where the blue distribution shows the simulated data and the black distribution shows the observed data. The values of spectral slopes have been computed in log-log space for simplicity. Red and black lines indicate the means of these distributions.}
  \label{fig:spectra}
\end{figure}

Here $m$ gives the spectral slope. This allowed us to get an estimate of the time evolution of the spectral slope during the flares and the spread of the values both in the observed and the simulated data. We constructed distributions of the spectral slope by first calculating the spectral slope for each simultaneous data point of the three energy bands of each sampled light curve, and finally, obtained the full distribution of a single simulation by combining the spectral slopes of each sample. These distributions were compared against the observed distribution using the two-sided Anderson-Darling test and the same criterion as before for accepting a match (see Sect. \ref{distr}). Figure~\ref{fig:spectra} shows an example of a comparison of the spectral slope distributions where the black histogram represents the observed data and the blue histogram represents the simulated data.

\section{Results}
\label{results}

In the following subsections, we describe the results of the tests introduced in the previous section, divided by the magnetic field of the simulations. As explained in Sect. \ref{setup}, we set up our simulations in a way that the jet power $P_{j}$ remains constant with varying magnetic field strengths $B$. Thus, the simulations with higher $B$ are designed to have shorter half-length of the layer $L$. The layer half-length directly relates to the observed duration of the reconnection event and the duration of the entire simulation. Some of the tests in our analysis rely on these data sets being approximately of the same size, and because of this, we needed to change certain details in the analysis process based on this division. These changes will be described before each section where necessary.

\subsection{Simulations with B = 0.1 G}
\label{b01sims}

Here we describe the results of our analysis for the simulations with the magnetic field strength of $B$ = 0.1 G. As explained in Sect. \ref{setup}, we chose to keep the jet power $P_j$, and according to Eq. (\ref{eq:jetpower}) the increase in the magnetic field strength has to decrease the dissipation distance $z_{diss}$ and thus the half-length of the reconnection layer $L$. Therefore, the simulations with $B = 0.1$ G have to have the longest $L$, and their observed duration spans approximately between 100 to 800 hours. Due to the duration and the dense temporal cadence of the observed data in comparison with the duration of the simulated data we are able to perform well-justified statistical comparisons using the methods described in Sect. \ref{methods}. Because the observed data spans across $\sim$200 hours, all the simulations longer than 300 hours were cut to 200-hour pieces before sampling them 1000 times as described in Sect. \ref{sample}. The simulations shorter than 300 hours were kept as they were and sampled 1000 times.

\begin{figure}
\centering
\includegraphics[scale=0.30]{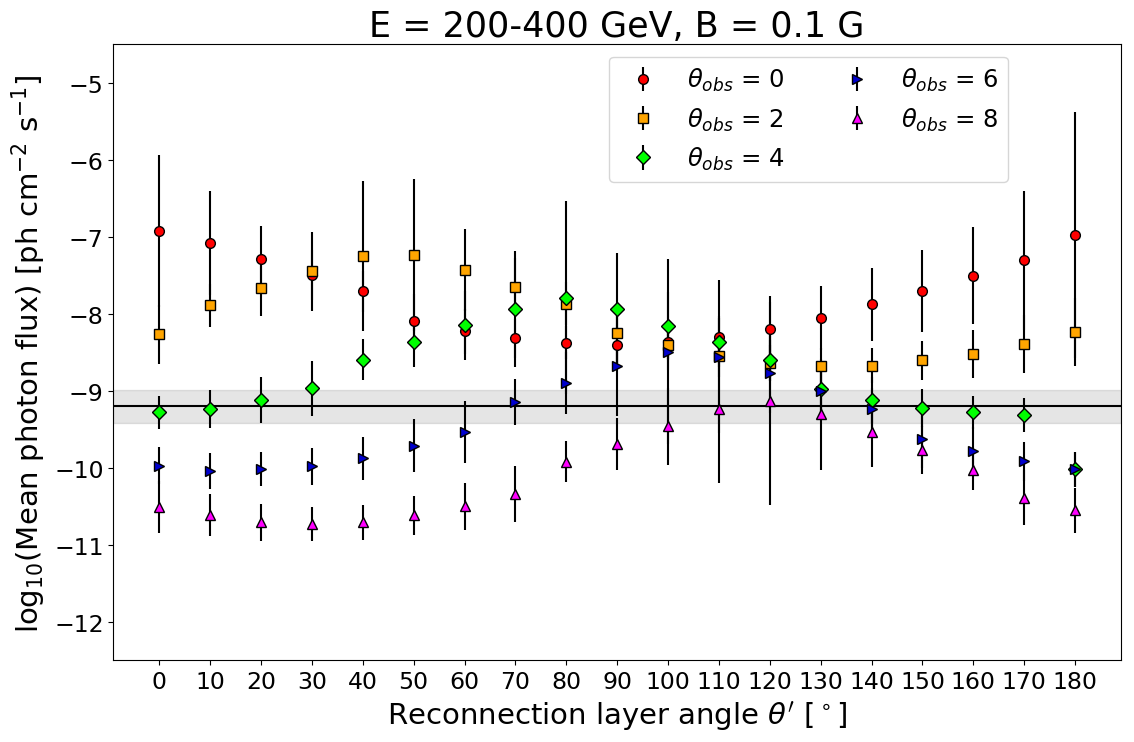}
\includegraphics[scale=0.30]{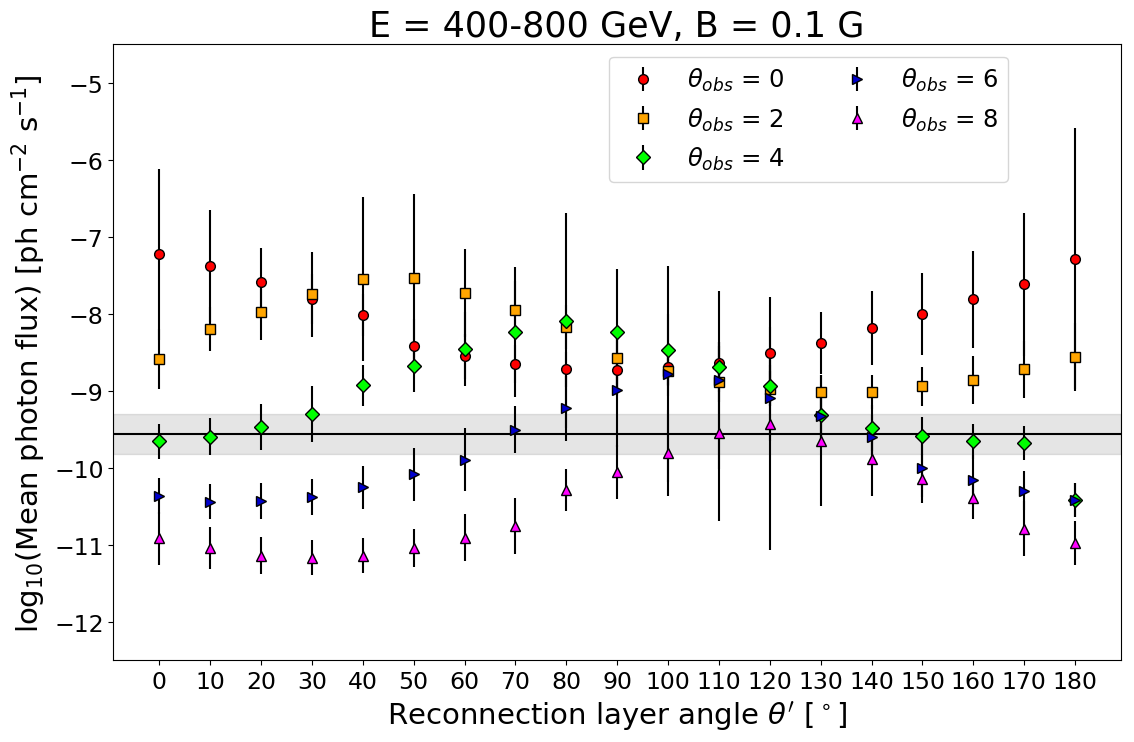}
\includegraphics[scale=0.30]{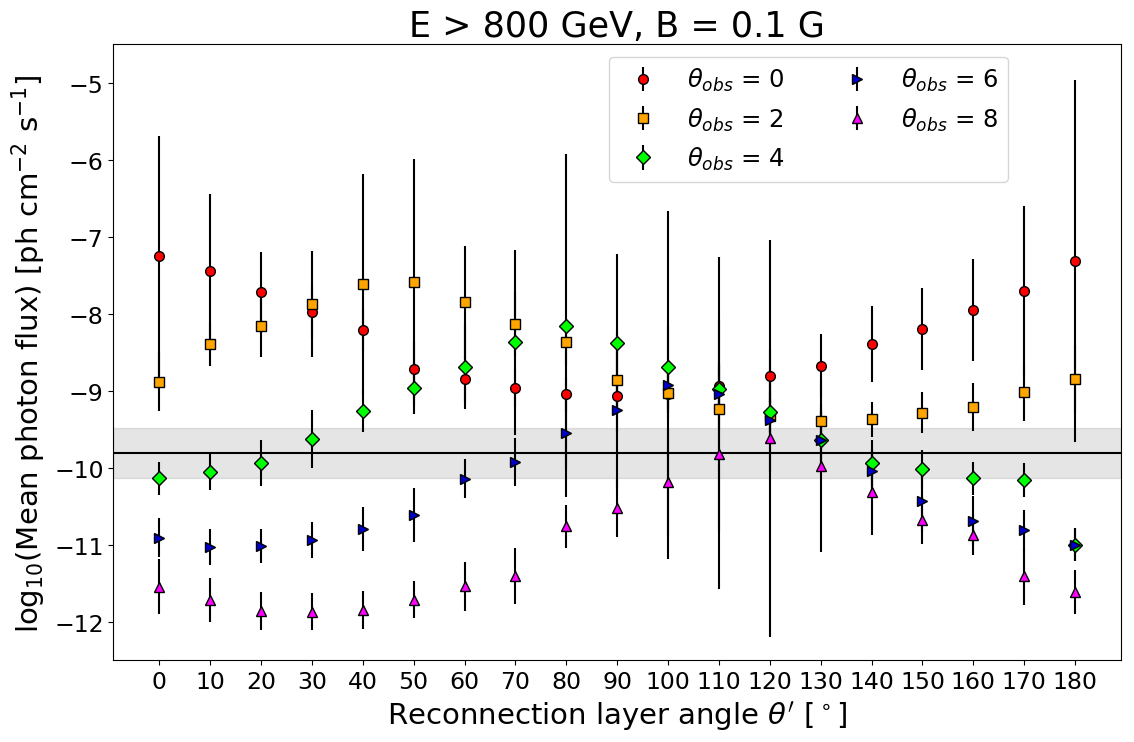}
  \caption{Mean fluxes of the simulations with $B$ = 0.1 G in each energy band, each colour representing a different viewing angle $\theta _{obs}$. The error bars represent the relative standard deviation. The black horizontal line with the shaded area shows the observed mean flux and the relative standard deviation.}
  \label{fig:meanfluxb01}
\end{figure}

In addition to the tests that were designed to give detailed information about the similarity of the data sets (see Sect. \ref{methods}), a simple estimation of the compatibility of our model was made by looking at the mean fluxes, namely we checked whether the simulated light curves are within the observable limits for this source in particular. However, as the simulations include several free parameters that can affect different attributes of the resulting light curve, we did not want to use the mean fluxes alone, or any of the single tests, to rule out an entire parameter space. The mean photon fluxes can be used to estimate the possible result of the flux distribution matches which require a close match of the general flux level as well as the overall shape of the light curve to resemble the observed data set. Figure \ref{fig:meanfluxb01} shows the mean flux of each viewing angle $\theta _{obs}$ and each energy band. This graphic shows that the reconnection layer angles resulting in the highest Doppler boosting, and in turn the highest fluxes, are shifting with each viewing angle. Moreover, the fluxes are lower in the higher energy bands because of the softer spectral shape in our model. For $\theta _{obs} = 4 \degree, \ 6\degree$, and $8 \degree$ we find simulations predicting fluxes within the observed range in all bands\footnote{For an animation of the time-evolved SED, as compared to the time-averaged observations from Mrk-421, for $\theta_{obs} = 8^\circ$ and $\theta^\prime = 100^\circ$, see \href{https://youtu.be/u77zEGqrkAs}{here}.}. These correspond to specific layer orientations $\theta ' = 0\degree -30 \degree, 140 \degree-180 \degree$ at $\theta _{obs} = 4 \degree$, $\theta ' = 60 \degree-80 \degree , 130 \degree -140 \degree$ at $\theta _{obs} = 6 \degree$, and $\theta ' = 110 \degree-130 \degree$ at $\theta_{obs} = 8 \degree$.

\begin{figure*}
\centering
\includegraphics[scale=0.30]{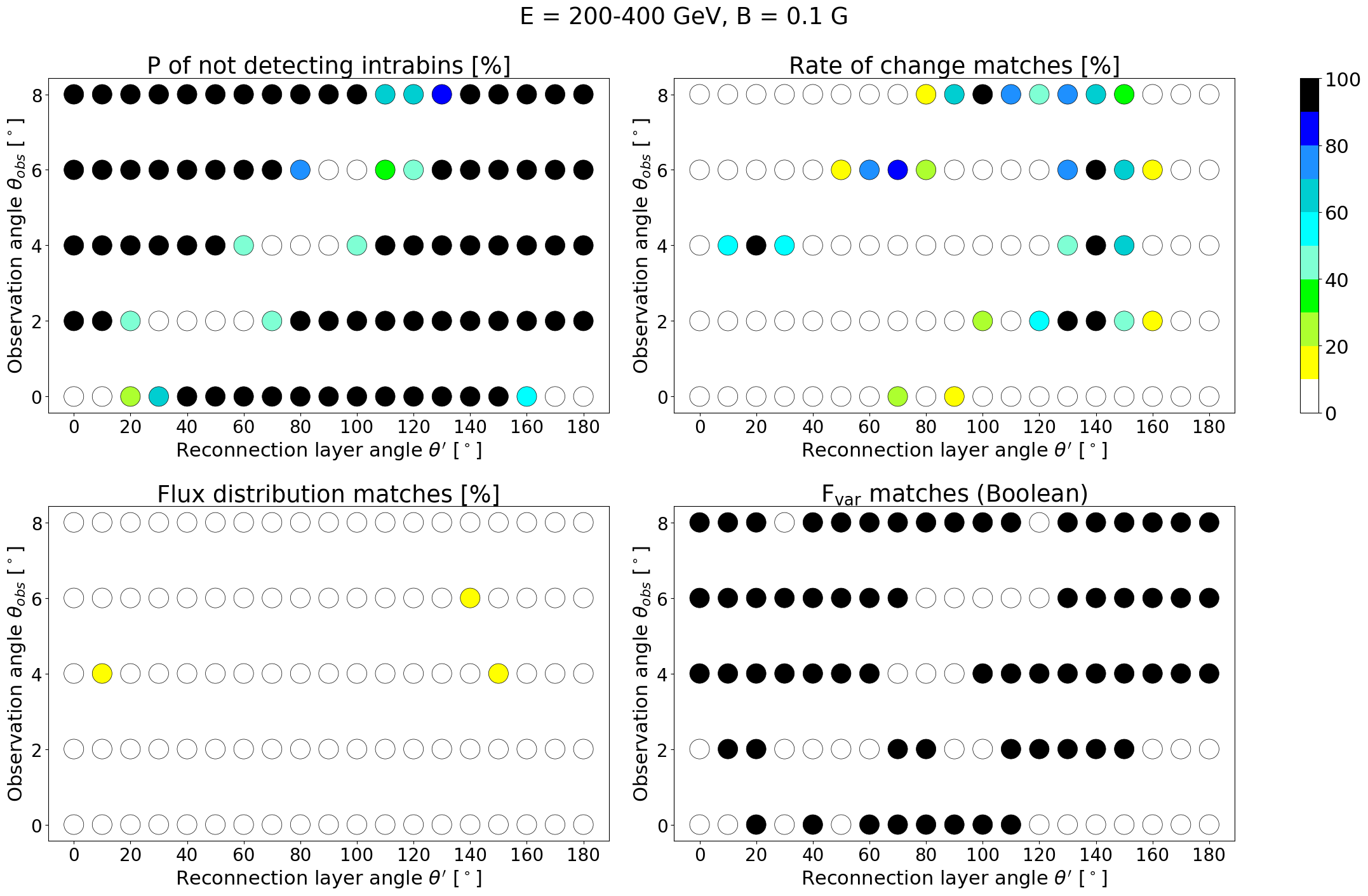}
  \caption{The results of the individual tests described in detail in Sect. \ref{methods}. The upper left panel shows the results of the intrabin test, describing the probability of not detecting intrabin variability in each of the simulated light curves. The upper right panel shows the results of the rate of change test, describing the fraction of matching simulated samples. The lower left panel shows the results of the flux distribution test, again showing the fraction of the matching simulated samples. The lower right panel describes the results of the fractional variability test, depicting the results in Boolean logic, with 100$\%$ signifying a match and 0$\%$ a non-match.}
  \label{fig:ind200}
\end{figure*}

As described in Sect. \ref{intrabin}, we calculate the probability of observing variability when the flux triples within shorter time scales than the observed 15-minute bins for all simulated light curves. Overall, the best candidates based on this test are those simulations where we are least likely to observe intrabin variability, but low flux is a limiting factor for observing the fastest variability time scales with current generation IACTs. The results of the individual tests are summarised in Fig. \ref{fig:ind200} for the energy band between 200 - 400 GeV and in Figs. \ref{fig:ind400} and \ref{fig:ind800} for the bands 400 - 800 GeV and >800 GeV respectively. The results of the intrabin test are shown in the upper left panel where we can see an abundance of simulations where we do not detect any intrabin variability.

A more detailed comparison of the time scales is done by comparing the rate of change of the simulated flares with those of the observed data. These results shown in the upper right panel of Fig. \ref{fig:ind200} for the 200 - 400 GeV band are given as a percentage of samples with matching flare rates of change out of a 1000 samples. The best matching simulations are those showing a high percentage of matches. In addition to this, the duration of each simulated light curve was calculated after the extraction of the low flux values (see Sect. \ref{tails}), leaving behind only the true reconnection event. The duration of the entire simulation also correlates with the time scales of the found flares, meaning that the fastest flares are found in the shortest simulations. The detailed results of the intrabin and the rate of change tests are collected in Table \ref{tab:B01_results2}. The calculated durations of the simulations are collected in Table \ref{tab:B01_results1}.

As explained in Sect. \ref{amp}, we searched for the matching distributions of flux, and we present the results as the percentage of samples showing matching distributions in the lower left panel of Fig. \ref{fig:ind200} for the 200 - 400 GeV band. Evident from these results, this test is very strict as there are only few simulations that show matching distributions in the two lower bands, the 200-400 GeV and the 400-800 GeV bands, and (almost) none that show matches in the highest band. The matches found, although few, occur in those simulations found favourable also by the time scale comparisons. The fact that we find none of these simulations matching in all energies would indicate that while many of the simulations match the observed flux levels, the combined effect of the shape of the light curve and spectrum does not match in any of the simulations. 

The fractional variability distribution matches are either matching: "yes" or non-matching: "no" (see Sect. \ref{fvar}). The results shown in the lower right panel of Fig. \ref{fig:ind200} for the 200 - 400 GeV band demonstrate that this test is not as strict and we can find several simulations where some or all of the bands have a distribution of fractional variability factors where the observed variability could possibly originate from. The results from both flux amplitude tests are in general agreement with the results of the time scale tests. The detailed results of the flux distribution and the fractional variability tests are collected in Table \ref{tab:B01_results3}.

\begin{figure*}
\centering
\includegraphics[scale=0.30]{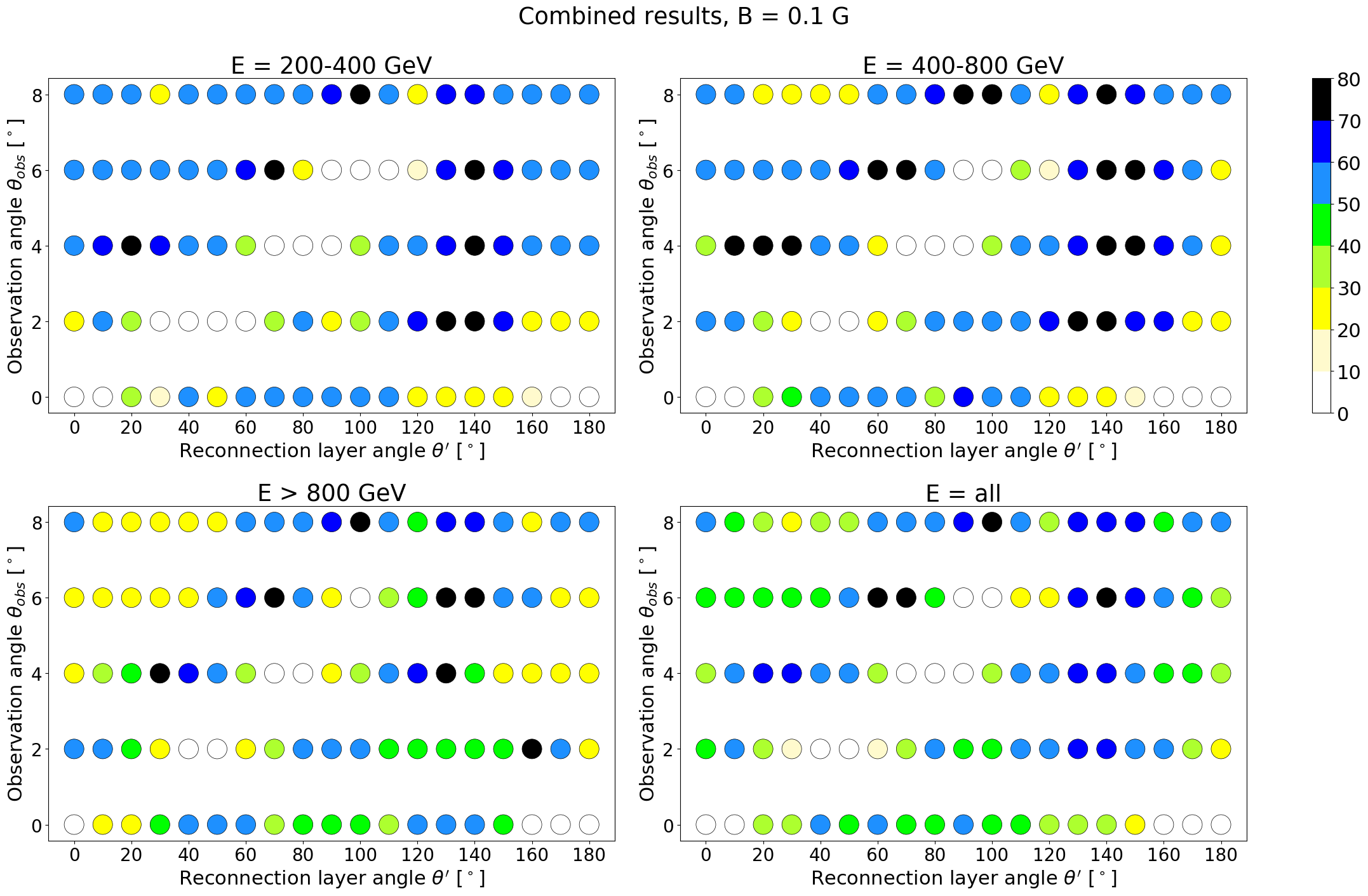}
  \caption{The combined test results for the simulations with $B = 0.1$ G for the three energy bands and also the results of all bands together (lower right panel). The individual test results shown in Figs. \ref{fig:ind200}, \ref{fig:ind400}, and \ref{fig:ind800} are combined first separately for each respective energy band. Each test has an equal contribution weight to the combined result, i.e. a 100\% match in an individual test means a 25\% match in the combined test of an individual energy band and an $\sim8.3\%$ match when combining all bands. The \textit{gold sample} is defined as those simulations that exceed the 70\% match threshold in the combination plot of all energy bands.}
  \label{fig:superresults}
\end{figure*}

The results of the individual tests are combined in Fig. \ref{fig:superresults} in such a way that each of the four tests has an equal weight and can contribute a maximum of 25\% to the combined result in a case where the individual test shows a 100\% match for the simulation in question. The lower right panel shows the combined results from all energy bands, thus adding the results of the first three panels together and normalising this to show an overall percentage. From this plot, we define the best matching sample of simulations, the \textit{gold sample}, as those that reach above the threshold of 70\% of a match. These simulations are namely $\theta _{obs} = 6\degree$ with $\theta' = 60\degree, \ 70\degree$ and $140\degree$, and $\theta_{obs} = 8\degree$ with $\theta' = 100\degree$.

In addition to the above four tests, the results of the spectral properties are shown in Table \ref{tab:B01_results4}. We calculated the spectral slope distribution matches as a percentage of matching samples for each simulation, but we find no matches for any of the simulations. Based on the flux distributions and this result, it is evident that the simulated spectrum does not describe the observed data particularly well in the highest energies of 800 GeV and above. As an indicative result, we calculate the mean of the spectral slope distribution of individual samples and take the mean of the obtained values to see how it compares with the observed mean. Although we find simulations in the viewing angle $\theta_{obs} = 8\degree$ with similar means, their standard deviations are found to be more narrow than those of the observed distribution. This is further elaborated in Sect. \ref{discussion}. 

\subsection{Simulations with B = 1 G}
\label{b1sims}

In this section, we describe the results of the simulations with the magnetic field strength of $B$ = 1 G and discuss the robustness of these results in comparison with the $B = 0.1$ G simulations. With the increased $B$, these simulations have shorter half-length of the layer $L$, and thus, they span between 10 to 80 hours. Because of this large variation in duration in comparison with the observed data set ($\sim$200h), we divided these simulations into the following three categories:
\begin{enumerate}
    \item Simulations shorter than 15 hours resemble approximately observational duration of a single night, and were therefore compared with those nightly light curves of Mrk 421 where there were enough data points for a statistically meaningful comparison, that is the sample size, number of data points in this case, had to be large enough. The first six nights of the observed data were deemed to fulfill these criteria. As a result, we have 1x6 comparisons.
    \item Simulations between 15 and 48 hours (i.e. approximately between one to two observational nights) were sampled using a sliding window down to 10 hours to resemble the observational duration of a single night. The window was shifted ten times to acquire meaningful variation between each sample. This results in 10x6 comparisons when comparing with the first six nights of the observed light curves.
    \item Finally, those simulations that were longer than 48 hours (i.e. longer than two observational nights) were sampled in a similar manner than the $B$ = 0.1 G simulations to include the daily gaps in the light curves but instead of a 1000 samples, they were sampled only 30 times to acquire reasonable variation between each sample. In addition, because the observed data set spans across ~200 hours, the observed light curves were sampled to similar lengths as each simulation longer than 48 hours an additional 30 times, resulting in 30x30 comparisons.
\end{enumerate}
This categorisation is necessary for the analysis of the shortest light curves of these simulations since their appearances are restricted by the temporal cadence of the observed light curve. This means that we are not able to compare the variations of the theoretical fluxes within a shorter temporal cadence than the chosen binning of the observed data (as explained in Sect. \ref{intrabin}) but neither can we generate more samples out of such short light curves. Therefore, we are forced to perform these analyses with the unequal sample sizes based on the above division, and it is important to keep in mind that unlike for the $B$ = 0.1 G simulations, we are not able to maintain a strong statistical consistency in the analysis of these simulations.

\begin{figure}
\centering
\includegraphics[scale=0.30]{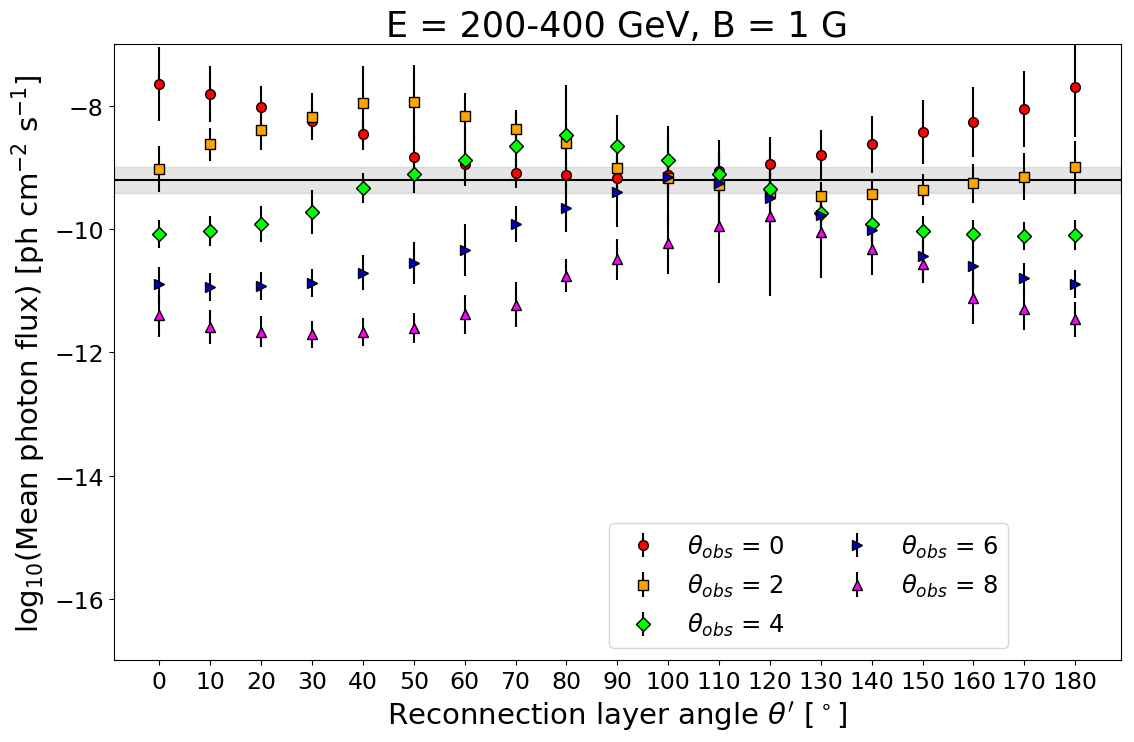}
\includegraphics[scale=0.30]{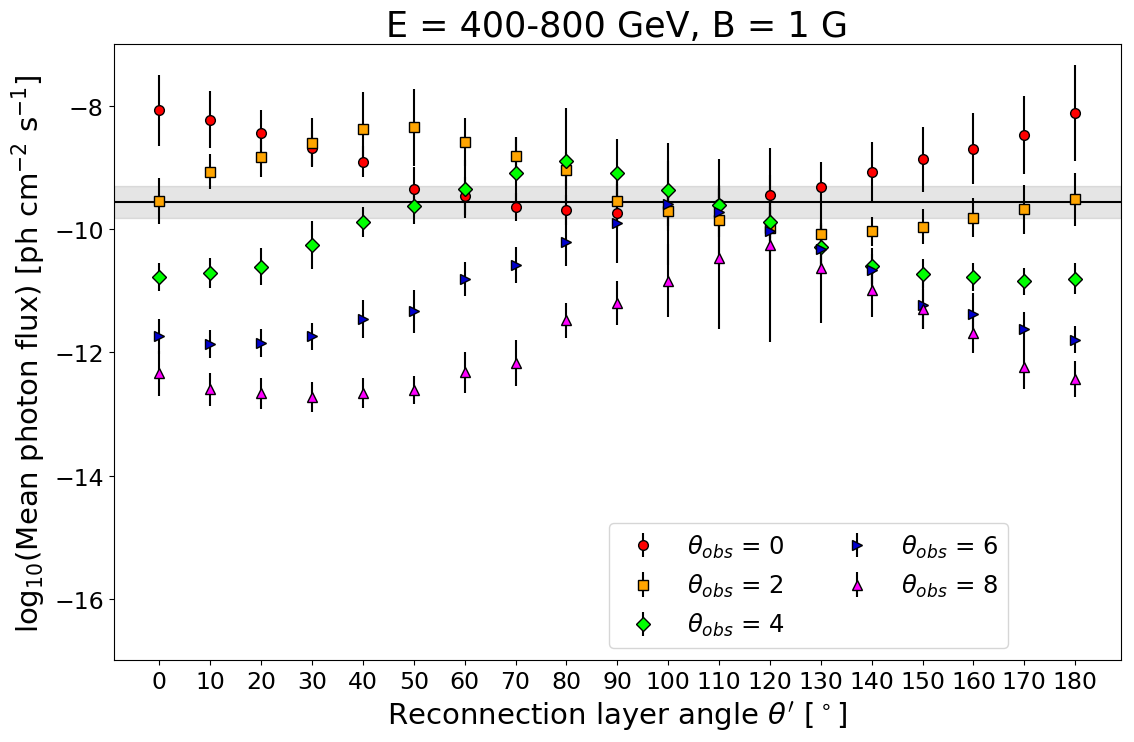}
\includegraphics[scale=0.30]{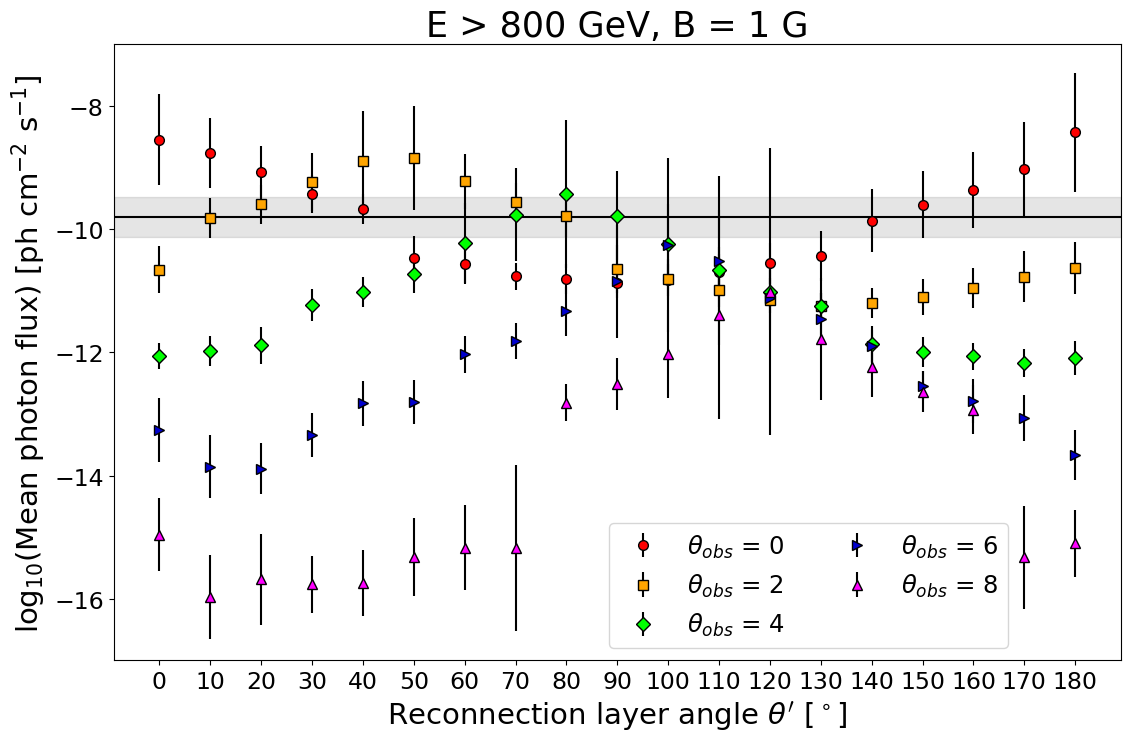}
  \caption{Mean fluxes of the simulations with $B$ = 1 G in each energy band, similar to Fig. \ref{fig:meanfluxb01}.}
  \label{fig:meanfluxb1}
\end{figure}

Figure \ref{fig:meanfluxb1} shows the mean fluxes and standard deviations of the simulations plotted together with the observed mean and standard deviation where we can clearly see that especially in the highest band $>$800 GeV many more simulations are falling below the observed mean. This also indicates that the shape of the spectra for these simulations is typically much softer compared to the observed spectrum and the spectra of the simulations with $B = 0.1$ G. This results from the decreased contribution of the high energy component of the SED due to the higher $B$. However, especially at $\theta _{obs} = 0 \degree$ with the highest Doppler boosting these fluxes are still at a rather similar level or higher than the observed fluxes. The durations and mean fluxes of these simulations are summarised in Table \ref{tab:B1_results1}.

The intrabin variability is calculated in a similar manner as for the simulations with B = 0.1 G. Because of the shorter duration of these simulations, they also possess more fast variability, and thus, more data points with intrabin variability time scales. However, there are still several favourable simulations where we do not detect any intrabin variability. Similarly, favourable layer angles $\theta '$ can be found for each $\theta _{obs}$ in terms of the flare rate of change, but these results cannot be expected to be as reliable since we are missing the more robust statistical dimension of this test in many cases where the duration of the simulation matches only one or two observational nights. The results of these tests are summarised in Table \ref{tab:B1_results2}.

In terms of flux distributions, the matches are found only in viewing angles $\theta _{obs} = 0 \degree, 2 \degree, 4 \degree$, and no matches are found for larger viewing angles due to lower fluxes. For the full fractional variability test described in Sect. \ref{fvar}, we do not have large enough sample sizes for these simulations, and therefore cannot construct a distribution of the fractional variability factors to perform a statistically meaningful comparison. In turn, we simply calculate the fractional variability factor and its error, and computed a zeta-score \citep{AN9952002303} for the comparison of two values within errors to assess the similarity of the fractional variability factors of these light curves. The zeta-score is defined as
\begin{align}
\zeta _i = \dfrac{X_i - X_{ref}}{\sqrt{u_i^2 + u_{ref}^2}},
\end{align}
where $X_{ref}$ is the reference value, in this case, the observed fractional variability factor, that the simulated value $X_i$ is being compared to, and $u_{ref}$ and $u_i$ are their respective error estimates. Fractional variability factors with $\zeta \leq 2$ are regarded as similar to each other. We find most matches of fractional variability in the larger viewing angles. However, this might be a result of a bias stemming from the longer observed duration of these simulations. The longer duration simulations naturally can have more samples and thus more possibilities of finding matches compared to the shorter simulations in the smaller viewing angles where we have fewer samples or only the single full light curve. The results of the flux amplitude tests are summarised in Table \ref{tab:B1_results3}.

As explained above, the mean fluxes depicted in Fig. \ref{fig:meanfluxb1} diverge from the observed value dramatically in the highest band, leading to softer spectra than in the observed data. We compared the simulated spectral slope distributions within the limitations of these simulations and find only two of the simulations with $\theta_{obs} = 0\degree$ matching with one night of the full observed light curve.

For clarity, all the results are presented as percentages, but because of the unequal number of samples, these percentages do not offer us an objective estimation of the compatibility of these simulations.

\subsection{Simulations with B = 10 G}
\label{b10sims}

\begin{figure}
\centering
\includegraphics[scale=0.30]{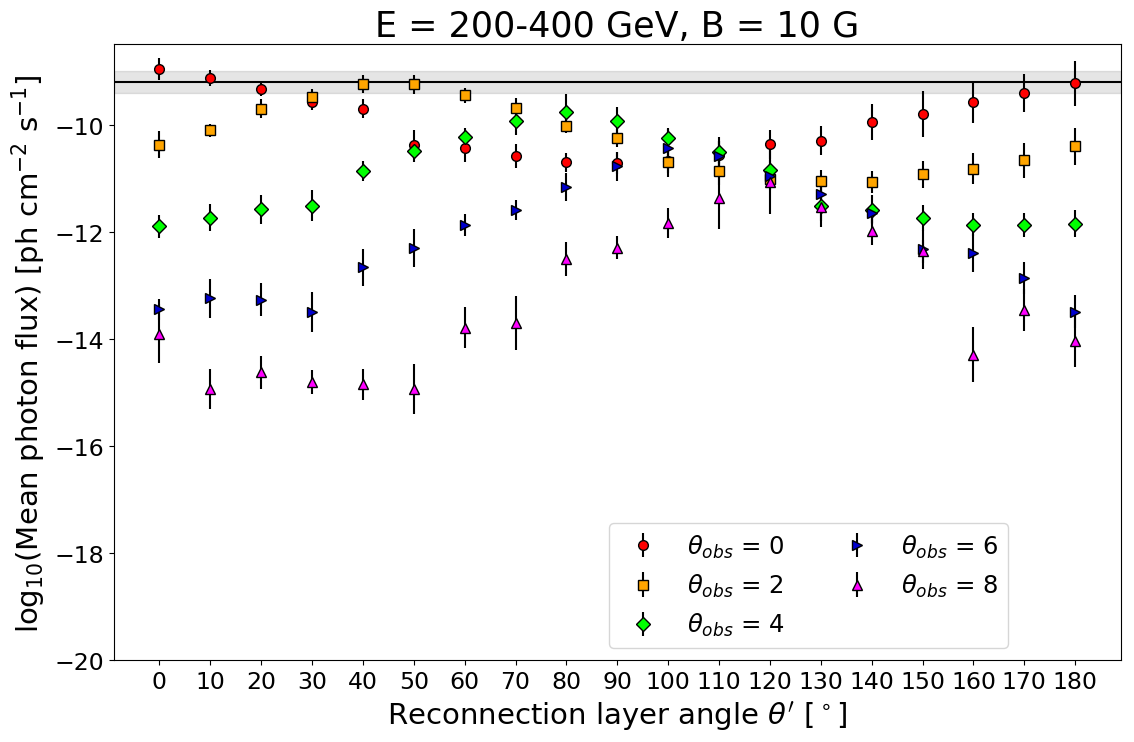}
\includegraphics[scale=0.30]{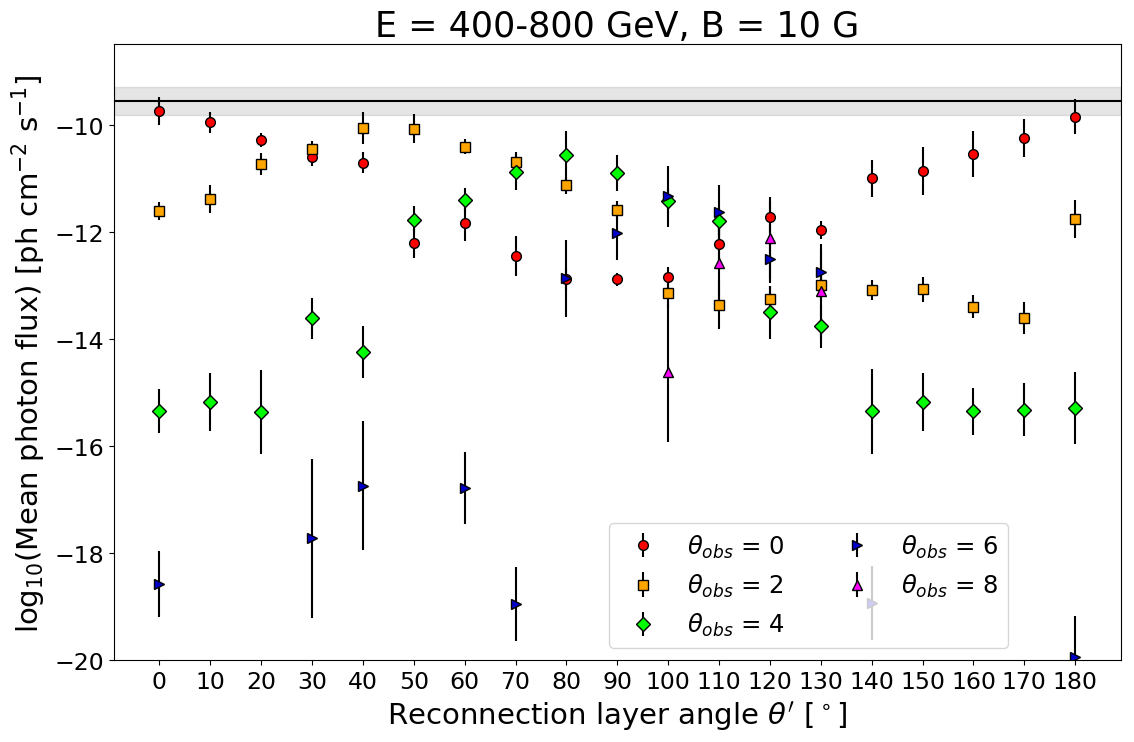}
\includegraphics[scale=0.30]{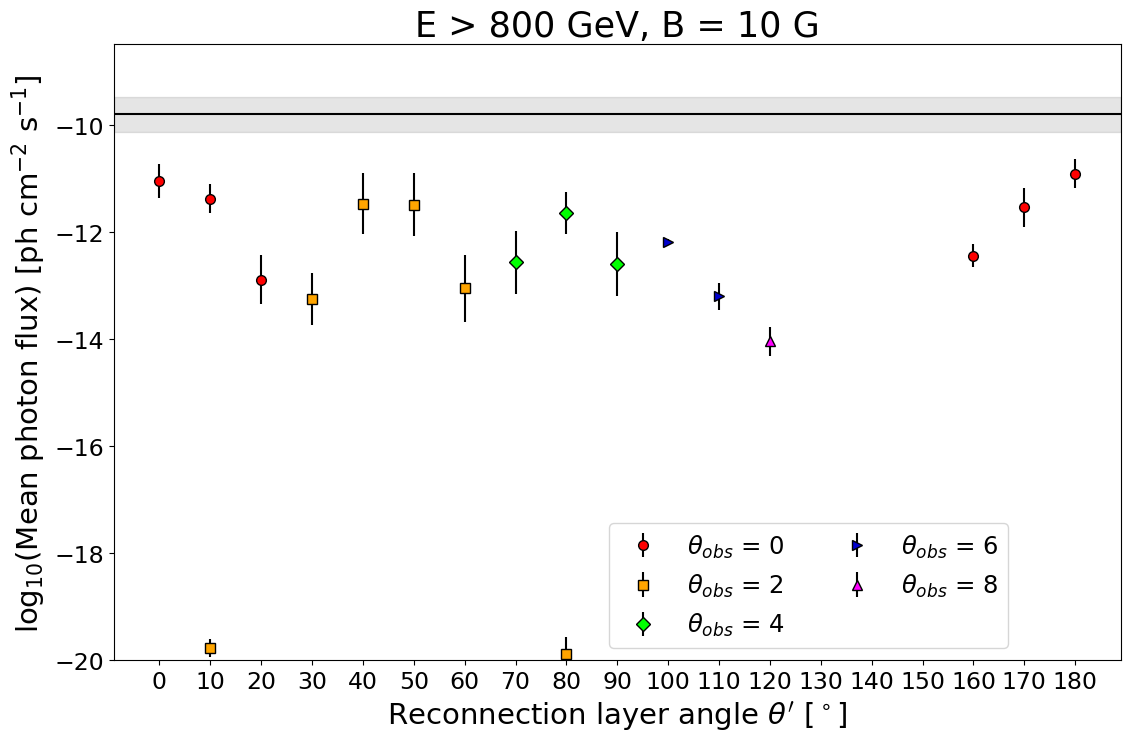}
  \caption{Mean fluxes of the simulations with $B$ = 10 G in each energy band, similar to Fig. \ref{fig:meanfluxb01}.}
  \label{fig:meanfluxb10}
\end{figure}

The final set of simulations that we considered for this source was that with the highest magnetic field strength, $B$ = 10 G. Because of this, however, the half-length of the layer for these simulations was the shortest and the simulation durations were cut down to 1 to 8 hours, as shown in Table \ref{tab:B10_results1}, thus leaving us with very few data points in many cases when using the 15-minute binning as per the source light curve. The mean fluxes of these simulations are shown in Fig. \ref{fig:meanfluxb10} in comparison with the observed mean. We can see that the simulated fluxes are significantly lower than the observed photon flux mean, and that in the highest band many of the simulations the fluxes are not within physically meaningful limits. In addition to this, we find that the variability in these simulations is in almost all cases much faster than the observed 15-minute binning (see Table \ref{tab:B10_results2}.  Both of these results were used to definitively rule these simulations out for this particular observed light curve.

\section{Discussion}
\label{discussion}

In this paper, we present the first broad and systematic comparison of a magnetic reconnection model for blazar emissions against observed data. Our approach is using the observed data in two different stages. First, in setting up the simulation parameters to ranges that best match the observed or theoretical values recorded for our source, Mrk 421, and second, in the comparison of the simulated light curves against the observed light curves. In order to narrow down the parameter space, we have developed a set of methods described in Sect. \ref{methods}, which we used in the assessment of the different features of these data sets. In this section, we discuss further the compatibility of our model to the observations, bring up caveats in our methods, and point out how this study can guide us forward in the exploration of magnetic reconnection models in application to VHE gamma-ray blazar flares.

The framework that we have used in this work to generate the theoretical light curves is unique in the sense that the derived light curves are based on a physical model for the plasmoid chain formed within a reconnection layer with implemented radiative properties. Therefore, we are able to compare the simulations with the observed data almost directly after the initial treatment of the theoretical light curves (see Sect. \ref{treatment}). Because of this possibility, in addition to being able to restrict the current model that we use, we were able to explore the capabilities of our model further by iterative work of careful tuning of parameters. In turn, despite the successful results of our analysis, our model has deficiencies that either need to be accounted for in future studies or, at least, one should be aware of when interpreting our results. As part of the analysis, we explored the capabilities of our model in several ways in order to improve the compatibility with the observed data, and in order to understand in which ways our theoretical model does not adequately represent the observed data.

\subsection{Significance of the results}
\label{disc:significance}

As can be seen in Fig. \ref{fig:ind200}, each of the developed tests has a different level of restricting the parameter spaces while still being in general agreement with each other. Because individual tests have caveats to act as a sole motive for discrimination, only by combining their results we can probe the most compatible models for this source in greater detail than ever before (see Fig. \ref{fig:superresults}). We have shown that with our methodology we are able to further constrain the already narrowed down, initial gamut of models, and we obtain the \textit{gold sample} of simulations that best match with the observed light curves\footnote{For animation of the time-evolved SED, as compared to the time-averaged observations from Mrk-421, for $\theta_{obs} = 8^\circ$ and $\theta^\prime = 100^\circ$, see \href{https://youtu.be/u77zEGqrkAs}{here}.}. These simulations with $B = 0.1$ G correspond to the viewing angle $\theta_{obs} = 6\degree$ with reconnection layer angles $\theta^\prime = 60 \degree, 70 \degree$ and $140\degree$, with maximum observed Doppler factors are $\sim 18$, $22$, and $9$, respectively, and $\theta_{obs} = 8\degree$ with reconnection layer angle $\theta^\prime = 100\degree$ with a maximum observed Doppler factor of $\sim 25$.

The results of our analysis are interesting because our best matching simulations are with larger viewing angles than what is usually assumed for the SED fitting, but still the maximum Doppler factors are in agreement with lower limits required to avoid gamma-gamma absorption \citep{Dondi1995,Acciari2020}. The discrepancy between the observed jet speeds and the observed luminosities and fast variability in the VHE gamma-rays is known as the "Doppler crisis" and is seen in many sources alike \citep[e.g.][]{Acciari2019}. This has been suggested to result from a structured jet with a fast spine and a slow sheath layer, as is observed through VLBI, where the regions responsible for the gamma-ray emission are thought to be included in the faster spine \citep{Ghisellini2005}. Our result favours a suggestion by \citet{Giannios2013a}, as even if the viewing angle of the jet is rather large, the maximum Doppler factors of the gamma-ray emitting regions can still be high as the reconnection layers are misaligned.

As explained in Sect. \ref{setup}, we used a higher value of the jet bulk Lorentz factor $\Gamma_j$ than what was initially selected based on the theoretical upper limit from VLBI observations (i.e. $\Gamma_j = 4$). This choice was based on our previous work, reported in \citet{Jormanainen2021}, where the lower $\Gamma_j$ and a shorter half-length of the reconnection layer $L$ yielded us significantly lower fluxes than in the observed data set of Mrk 421. For the simulations presented in this work, the initial values of $\Gamma _j$ and $L$ were multiplied by 3 to obtain simulated fluxes in the observed range (see Figs. \ref{fig:meanfluxb01} and \ref{fig:meanfluxb1}). While the layer length would still lie within physically reasonable limits, the bulk Lorentz factor for this source does not correspond to the value derived from observations. The maximum Doppler factors of the plasmoids in each of our \textit{gold sample} scenarios are mostly higher than those derived from radio observations (see Appendix~\ref{dopp_golden} and Fig.~\ref{fig:golden_dop}). This offers us an explanation for the discrepancy between the VLBI and VHE observations. In our \textit{gold sample} simulations, the large viewing angles of the jet would result in the slow or non-moving jet speeds observed in the VLBI but the misaligned reconnection layer orientations yield the high Doppler factors required to produce the high and variable emission we observe from these sources. This result offers us a plausible explanation for the long-standing problem in the study of blazar jets and is a major argument for the relativistic magnetic reconnection as an acceleration mechanism in the VHE gamma-ray regime.

Our analysis also provides a method to determine the viewing angle of the jet. For high synchrotron peaked objects like Mrk 421 it is challenging to estimate the viewing angle based on VLBI observations because the subluminal apparent component speeds do not necessarily describe the underlying speed of the flow \citep{lico12}. Even for the largest radio flare of Mrk 421 observed in 2012, the highest radio Doppler factor obtained was between 3 and 10 \citep{hovatta15}, and no new superluminal components were detected \citep{richards13}. Based on MOJAVE data \citep{lister11, homan21}, the high synchrotron peaked sources are distinguished from the rest of the blazar population by lower than average radio core brightness temperatures, lack of large-amplitude radio flares, and low linear core polarisation levels. All this would indicate that these sources possess rather low Doppler boosting and therefore rather large viewing angles and our findings are in agreement with that.

\subsection{Caveats of the model}
\label{disc:modelcaveats}

In this work, we benchmark our model with the observed data already prior to generating the simulated light curves (see Sect. \ref{setup}) in order to produce results that match the conditions of our observed source, Mrk 421, as close as possible. However, some caveats remain: jet energy density pre-factor, softer spectral shape, and a narrow distribution of spectral indices.

In Sect. \ref{setup}, we already discussed that the common feature of the magnetic reconnection models is to assume the plasmoids to be characterised by rough equipartition between magnetic fields and relativistic particles, which typically gives Compton ratios much lower than unity. However, during flaring, the SEDs of high synchrotron peaking BL Lac blazars such as Mrk 421 can show a stronger IC component with Compton ratios close to unity. \citet{Christie2020} suggested that the larger plasmoids could provide the necessary seed photon field to be upscattered through the IC mechanism by the smaller plasmoids. The energy density estimated for a large plasmoid as $7 \cdot 10^{-4} \ \mathrm{erg/cm^2}$ in its own frame can be close to $10-30$ times larger when measured in the frame of a smaller plasmoid due to the relative motion between the plasmoids. Given that $U'_B$ is $\sim 8\cdot 10^{-4} \mathrm{erg/cm^2}$, this additional factor of $10-30$ results in $U'_B/U'_e \sim 0.03 - 0.1$.

When we calculate the fraction of the energy densities for all plasmoids in our simulation in the observer's frame, we obtain a range of $U'_B/U'_e \sim 0.015 - 0.15$, consistent with the above estimate. Therefore, when measured in the frame of the observer, the inter-plasmoid scatterings would be enough to supply the seed photons in our model, which we here have corrected with the empirically searched pre-factors (see the discussion in Sect. \ref{setup} and Table \ref{tab:theory_params}). The use of scaled energy density fraction results in higher jet powers than the estimation used in our study, but as explained earlier, we can identify sources of uncertainty in the jet power acquisition methods and the exact jet power during the flaring epoch that would further justify the use of such a pre-factor. By combining these estimates of uncertainty, we reach at least a factor of $\sim 80$ increase to the jet power we have adopted.

As described in the previous section, our current set simulations were generated with a $\Gamma_j$ value (i.e. $12$) larger than that derived from VLBI observations (i.e. $\Gamma_j \approx 4$). Doing so, permitted the simulated fluxes to reach a level that is comparable to observations (see Fig.~\ref{fig:meanfluxb01}). However, our model shows a discrepancy in the number of matches and flux distributions between the two lower frequency bands (i.e. $200-400$~GeV and $400-800$~GeV) and the highest band (i.e. $> 800$~GeV). This mismatch implies that the model spectra do not fully describe the VHE observations. As an attempt to harden the simulated spectrum, the peak energy of the high-energy component of the SED was shifted by changing the maximum electron energy $\gamma_{max}$ within plasmoids for a particular configuration, namely $B = 0.1$ G, $\theta _{obs} = 4\degree$, and $\theta ' = 10\degree$, and $\Gamma _j = 12$. As a result, the observed spectrum with the updated $\gamma_{max}$ was much harder than the observed one. However, matching the observed spectrum becomes a much less trivial problem when looking at the time evolution of the SED for this source where high-energy SED peak can shift within days to as high energies as 1 TeV as shown in \cite{Acciari2019}. 

As is evident from Fig. \ref{fig:spectra}, the observed range of the spectral indices has a much wider spread than the simulated spectral indices. Naturally, some spread could also be introduced by adding noise to the simulated signal, but even with an added Gaussian noise we did not manage to acquire enough spread for the simulated spectral slope distributions. In terms of the adopted theoretical model, it is important to note that the slope of the injected particle spectrum was kept constant for all plasmoids to a value estimated by \cite{Sironi2015}. Although this assumption is approximate, it has been shown the slope of the particle spectrum within a plasmoid varies with time as the plasmoid traverses through the reconnection layer \citep{Hakobyan2021}, which can account for some but not all of the observed spread. Additionally, we assume that the observed flaring event is produced from a single reconnection event with magnetisation $50$, corresponding to injected particle power-law distributions of index $p\sim 1.5$ (while $p\sim 2$ for $\sigma = 10$). If instead, multiple reconnection events occur simultaneously in regions of different magnetisation (see, e.g. the striped-jet models of \citealt{giannios2019, Zhang2021} or jet-in-jet models \citealt{2010MNRAS.402.1649G}), one could expect a larger spread in the model spectra.

We noticed in our earlier work \citep[][where lower $\Gamma_j$ and $L$ were used]{Jormanainen2021} that some simulations have a tendency of showing extremely fast variability (see Sect. \ref{intrabin}). In order to reduce this, we ran a subset of seven simulations with the magnetisation $\sigma = 10$ (as opposed to the currently used $\sigma = 50$) and managed to lower the fraction of intravariable data points within these light curves. This also resulted in longer observed duration of the simulations in general. However, because with the current simulations (with a higher $\Gamma_j$ and a longer $L$ to increase the flux level) we still find an abundance of simulations with matching time scales, we did not pursue changing $\sigma$ for further tests. In the future, the simulations with higher magnetic fields ($B = 1, 10$ G), where we are not currently able to apply the statistical component of our analysis, could benefit from a lower value of $\sigma$ to increase their observed duration.

\subsection{Caveats of the methods}
\label{disc:methodcaveats}

The comparison of the simulated data to the observations is always limited by not only the capabilities of the model but also the quality and extension of the observed data. In this work, we have aimed to carefully consider the properties of the observed data in the treatment of the simulated data, but also in this sense there are still some caveats: shorter duration of higher magnetic field simulations where we cannot include a complete statistical analysis for them, choice of duration of the sampled light curves, and the flux amplitude tests not accounting for non-physical flare decay profile.

As discussed earlier, the problem of the short duration of the simulations limits our possibility to compare these with the observed data in a statistical manner. However, this does not mean that the duration in itself would intrinsically rule out the higher magnetic field strengths since flaring could result from several short reconnection events. We are missing the strong statistical dimension in the analysis of simulations with $B = 1$ G, but we could still consider some of these simulations to be a contributing part of our observed light curve, such as those with $\theta _{obs} = 0\degree$ that seem to agree with some of our tests. Ultimately, the statistical robustness of this analysis depends to an extent on the choice of duration of the simulated data and the binning of the observed data, that is how many data points construct our observed event and how much data the simulated light curve contains in comparison. For example, we have chosen to compare simulations with a duration between 100-300 hours with the observed data spanning 200 hours. We therefore also tested the effect of this choice by sampling a simulation that was sampled first as 300 hours to see if the matches persist when sampled down to 200 hours. For this one test case, our analysis results remained largely the same, but this cannot be said to hold in cases where we might have significantly less data. Even faster variability than the time scales observed for Mrk 421 has been observed for other sources \citep[such as the minute time scales seen in the radio galaxy IC 310][]{Aleksic2014}. Therefore, it is possible that the higher magnetic field strength could be applicable to these sources, and with a high temporal cadence better statistical comparison could be performed also with shorter time scale variability.

The restrictive nature of the flux distribution test relates closely to the shape and temporal evolution of the SED. As discussed in Sect. \ref{b01sims}, especially in the $>800$ GeV band we do not find any matches in almost all of the simulations, but also in the two lower bands, the matches reach 10-20\% at best. Matching the overall shape of the light curve is a difficult task because there are many things that affect the outcome of this test. In addition to the aforementioned SED shape and the general flux level, another aspect where our model is not matching the observed signatures is the decays of the flares (see Sect. \ref{tails} for explanation) that we disregard in the comparison of the flare rise times. But in the comparison of the flux distributions we are looking at the whole light curve, therefore including the flare decays in this test. The non-physical flare decay profiles are a caveat in our model but also in the tests of flux amplitudes since they do not account for this effect. In longer-term variability, largely symmetric flares are expected to result from the light-crossing time of the emission region \citep{Chatterjee2012}, but in a scenario where the fast flaring is caused by the acceleration of the particles, this might not be the case. However, because the flare decays in our model are not based on known physics of the fast flaring, we tested the effect of symmetric flares on our analysis. This was done by manually adding symmetric flare decays on one of the \textit{gold sample} light curves and rerunning the flux distribution, the fractional variability analyses, and the spectral slope distribution analyses. The symmetric flares increased the total flux and hardened the spectra of the tested theoretical light curves. These and the change in the shape of the overall light curve affected the found matches in both the flux distributions and the fractional variabilities. Mostly the effect was that those simulations where we had previously found matches in the flux distributions, no longer showed matches, but in one tested case, there was a slight increase of matches found in the highest energy band and additional matches in the middle band. All of the tested simulations showed fewer matches in the fractional variability test, but it could be expected that with the symmetric flares different simulations previously not matching as well in these tests might match better in a scenario where the simulated flares were symmetric. In future work, testing light curves with different decay profiles should be considered.

\subsection{Comparison with previous studies}
\label{disc:comparison}

Although this is the first time such an extensive comparison of simulated and observed data has been performed, models of relativistic magnetic reconnection have been applied to $\gamma$-ray blazar observations in the past. The detectability of plasmoid-powered $\gamma$-ray flares by the \textit{Fermi} Large Area Telescope (\textit{Fermi}-LAT) was first investigated by \cite{meyer2021}. These authors used similar theoretical models from \cite{Christie2019} to generate artificial $\gamma$-ray light curves to be compared with those observed from flat spectrum radio quasars (FSRQs). In general, the authors found that misaligned reconnection layers (i.e. $\theta^\prime \approx 30^{\rm o}$) produced light curves which have similar flux levels, variability, and power spectral density profiles as characteristic FSRQs (i.e. 3C 279 and 3C 273). In general, the binning applied to the artificial light curves washed out the variability on short timescales. Nonetheless, \textit{Fermi}-LAT would be able to detect plasmoid-powered minute-scale flares, as the one observed in 3C 273, if these coincided with the times when the source was in the field of view of the LAT.

Furthermore, magnetic reconnection was suggested as a means to produce the observed variability properties of Mrk 421 (i.e. the flux-doubling time scale and the peak bolometric luminosity) during the flare occurring on the night of April 15, 2013. In \cite{Acciari2020}, the authors adopted a simplified model for plasmoid emission, presented in \cite{Petropoulou2016}, which approximates the peak luminosity and flux-doubling timescale produced from a single plasmoid. By exploring a variety of different free parameters (e.g. $\sigma$, $\theta_{obs}$, $\theta^\prime$, etc.), the authors found that matching the luminosity and time scale required both a misaligned jet and reconnection layer, that is $\theta_{obs} \approx 2^\circ$ and $\theta^\prime = 30 - 90^\circ$. These results differ from our findings, as the range of $\theta^\prime$ which best matches observations for $\theta_{obs} = 2^\circ$ is determined to be $130-140^\circ$. The theoretical light curves used in this work are computed using the plasmoid motion dynamics in the reconnection layer as dictated by PIC simulations, while \cite{Acciari2020} used approximate relations. Additionally, \cite{Acciari2020} chose a magnetisation value of $10$, differing from our choice of $50$, which affects the terminal four-velocity of plasmoids in the layer (hence, the degree of beaming) and the injection spectrum of electrons in each plasmoid. As such, using PIC-simulated models could result in a more refined interpretation of the observed data.

\section{Summary and conclusions}
\label{concl}

This paper presents the first comprehensive scan of relativistic magnetic reconnection models compared against observations of fast VHE gamma-ray blazar flares in a quantitative manner. To simulate the light curves induced by magnetic reconnection, we used the model presented in \cite{Christie2019}. The model uses the results of state-of-the-art 2D PIC simulations for the sizes and velocities of plasmoids \citep{Sironi2016} coupled with a leptonic radiative model to describe the evolution of the radiating particles within a single plasmoid \citep{Petropoulou2016} and a leptonic radiative transfer model to track the evolution of the particle and photon spectra within each plasmoid \citep{Christie2019}. Prior to our analysis, we adjusted our simulation parameters specifically for the source, Mrk 421, with estimations of jet power, bulk Lorentz factor, peak frequency of the synchrotron spectrum, magnetisation of the jet plasma, and minimum Lorentz factor of the injected particle distribution obtained from the literature or through experimentation. With set ranges of magnetic field strength, viewing angle, and reconnection layer angle, we obtained a gamut of models describing different jet scenarios. The theoretical light curves were treated to resemble realistic observations of VHE gamma-ray data.

We developed a set of methods used to statistically study the compatibility of a theoretical model with real light curve data. In our analysis, we compared several aspects of the simulated and observed light curves. We first focused on the estimation of the time scales extracted from these light curves via tests of intrabin variability and rate of change of the detected flares. Next, we studied the flux amplitudes via tests of flux distributions and fractional variability. Finally, by utilising the observed data obtained in three energy bands, we performed a crude estimation of the spectral slope distributions. By combining the results of these different methods, we find the \textit{gold sample} of parameter spaces, namely simulations with $B = 0.1$ G, $\theta_{obs} = 6\degree$ with $\theta ' = 60\degree, \ 70\degree$ and $140\degree$, with corresponding maximum Doppler factors $\sim 18$, $22$, and $9$, and $\theta_{obs} = 8\degree$ with $\theta ' = 100\degree$, with maximum Doppler factor $\sim 25$ that best match the observed data. With these results, we demonstrate that we are able to limit the initial parameter space by using this variety of statistical tools. The simulations favoured by our model possess high Doppler factors, offering us an explanation for the Doppler crisis and a strong argument for the relativistic magnetic reconnection as an acceleration mechanism in the VHE gamma-rays.

In the future, we will apply these methods to other sources, such as the well-known TeV blazars BL Lac \citep{Abeysekara2018} and PKS 2155+304 \citep{Aharonian2007}, and the radio galaxy IC 301 \citep{Aleksic2014}, where intra-night variability has been observed in the VHE gamma-rays. The comparison with X-ray observations would be especially desirable in more careful determining of the high-energy SED component. With the methods developed for this study, it is possible to further explore the capabilities of our reconnection model by means of careful iterative tuning of parameters. Although we have focused on the comparison of relativistic magnetic reconnection models with the VHE data, our methodology is designed to be applicable to observations in other wavelengths as well as other emerging models of variability that can produce physically motivated light curves.

\begin{acknowledgements}
We thank the anonymous referee whose insightful comments helped us improve this manuscript. We thank Dr. Julian Sitarek for the helpful discussion on the determination of the sensitivity of MAGIC, and Dr. Cosimo Nigro for providing us with the script for the calculation of MAGIC sensitivity. We thank Dr. Axel Arbet-Engels for the helpful discussion on the SEDs of our observed data set. This work has made use of the published light curves of Mrk 421 observed by MAGIC and VERITAS, and we thank Dr. Ana Babić for providing the data and additional help. J. J. was supported by the Academy of Finland projects  320085 and 345899. T. H. was supported by the Academy of Finland projects 317383, 320085, 322535, and 345899. E. L. was supported by the Academy of Finland projects 317636 and 320045. M.P. acknowledges support from the MERAC Foundation through the project THRILL and from the Hellenic Foundation for Research and Innovation (H.F.R.I.) under the ``2nd call for H.F.R.I. Research Projects to support Faculty members and Researchers'' through the project UNTRAPHOB (Project ID 3013). I.M.C. and M.P. acknowledge support from the Fermi Guest Investigation grants 80NSSC18K1745 and 80NSSC20K0213.
\end{acknowledgements}

% WARNING
%-------------------------------------------------------------------
% Please note that we have included the references to the file aa.dem in
% order to compile it, but we ask you to:
%
% - use BibTeX with the regular commands:
%   \bibliographystyle{aa} % style aa.bst
%   \bibliography{Yourfile} % your references Yourfile.bib
%
% - join the .bib files when you upload your source files
%-------------------------------------------------------------------

\bibliographystyle{aa} % style aa.bst
\bibliography{vhe_paper} % your references Yourfile.bib

\begin{appendix} %First appendix

\section{\textit{Gold sample} Doppler Factors}
\label{dopp_golden}

\begin{figure*}[h]
\centering
\includegraphics[scale=0.45]{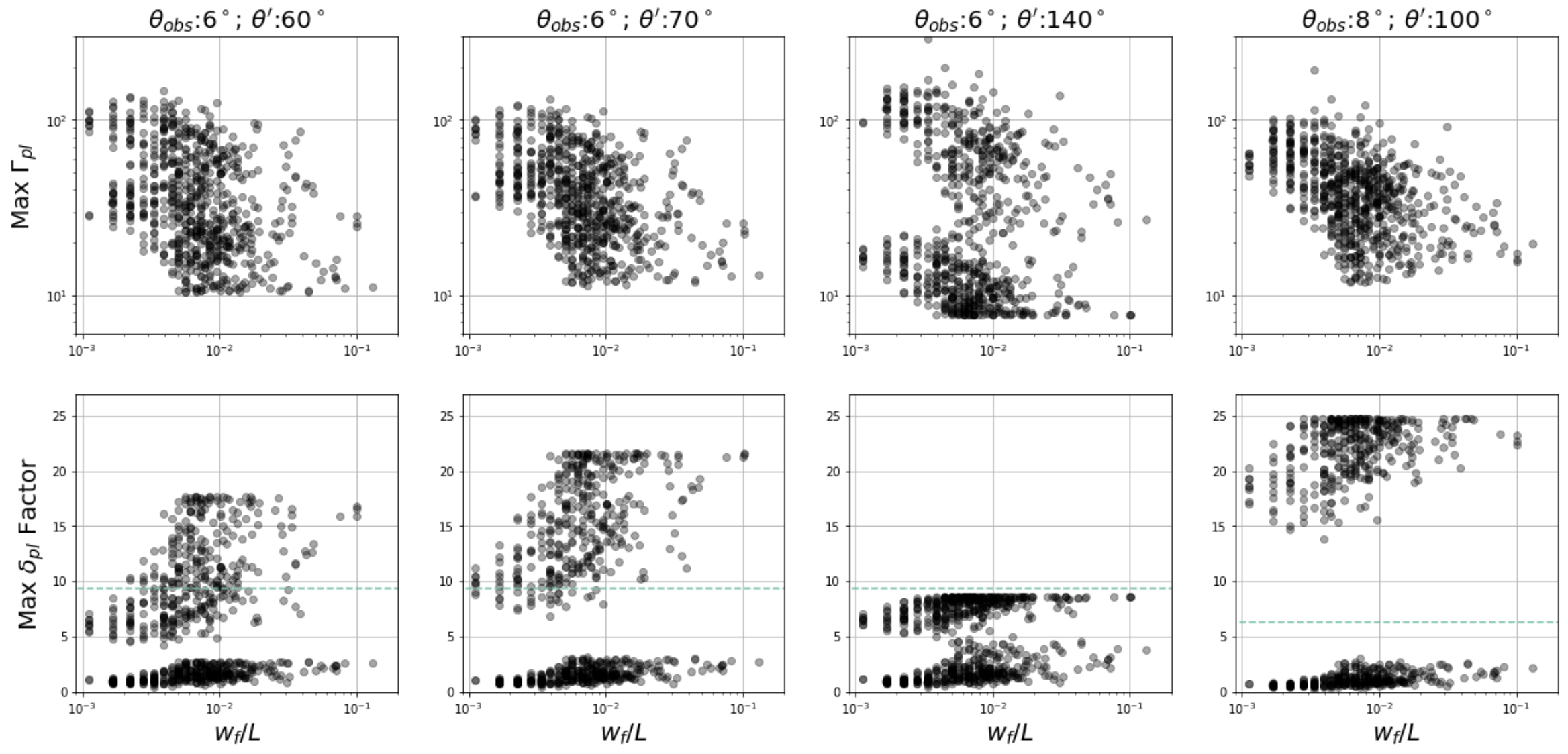}
  \caption{Max Lorentz factor (top row) and Doppler factor for all plasmoids as a function of final plasmoid size (normalised to the reconnection layer's half-length) for each of the four \textit{gold sample} orientations. The dashed green line within the bottom row denotes the observed Doppler factor of the jet.}
  \label{fig:golden_dop}
\end{figure*}

For some of the \textit{gold sample} simulations, the observed Doppler factors of the plasmoids, shown as a function of final plasmoid size in Fig.~\ref{fig:golden_dop}, are higher than those derived from radio observations. Additionally, for most of the orientations, the maximum Doppler factors are larger than the jet's Doppler factor (denoted by the dashed, green line in the figure).

\section{Individual test results}
\label{appa}

In this paper, we describe the methodology developed for the comparison of simulated light curves based on theoretical models against observed data. In order to obtain a comprehensive view of the different features of the two data sets, we have combined the results of several different tests in our analysis. Because our methods focus on different aspects of these light curves, we want to emphasise the importance of each of the individual tests by showing also their results separately.

\begin{figure*}
\centering
\includegraphics[scale=0.30]{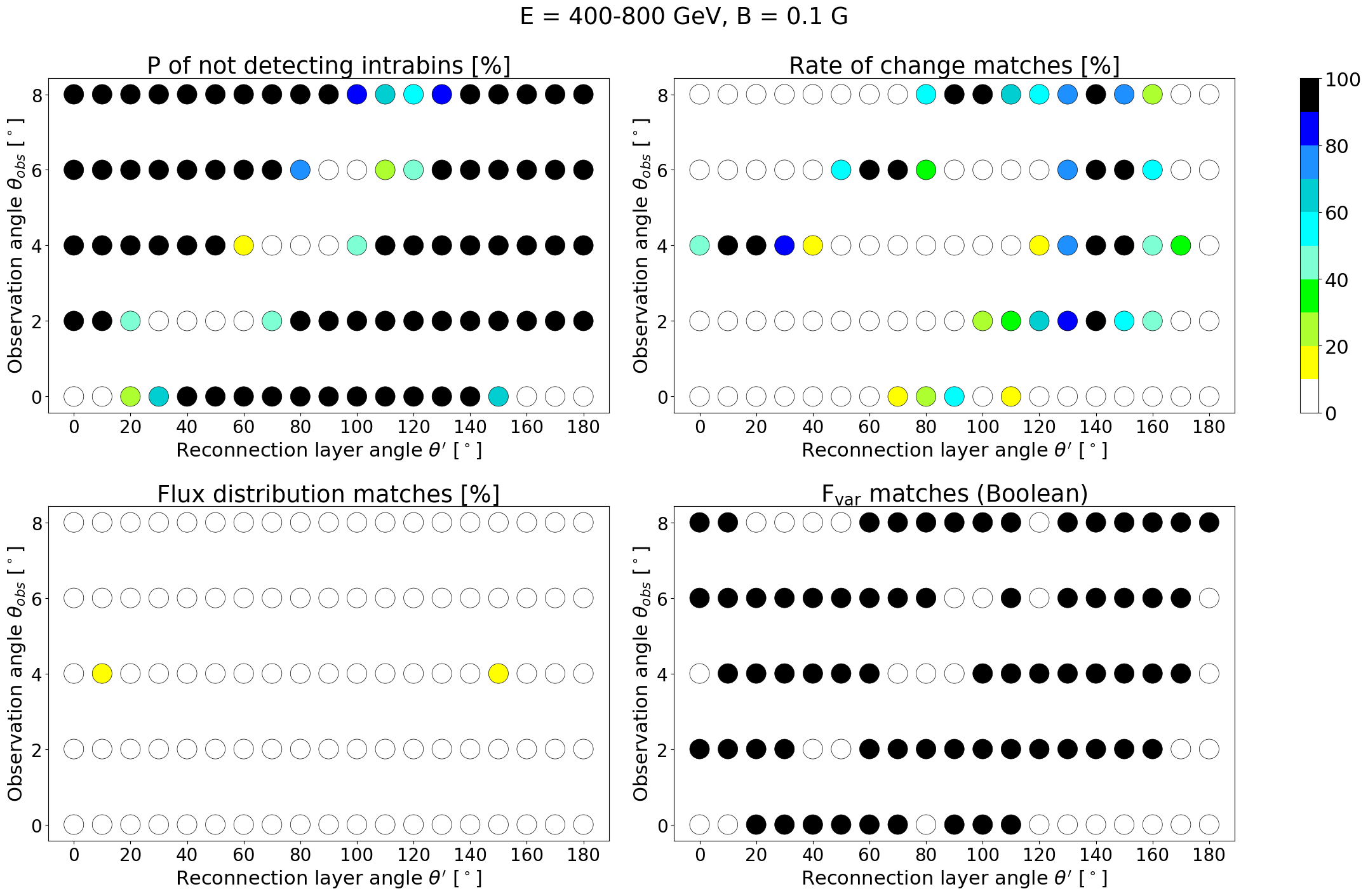}
  \caption{See description in Fig. \ref{fig:ind200}.}
  \label{fig:ind400}
\end{figure*}

\begin{figure*}
\centering
\includegraphics[scale=0.30]{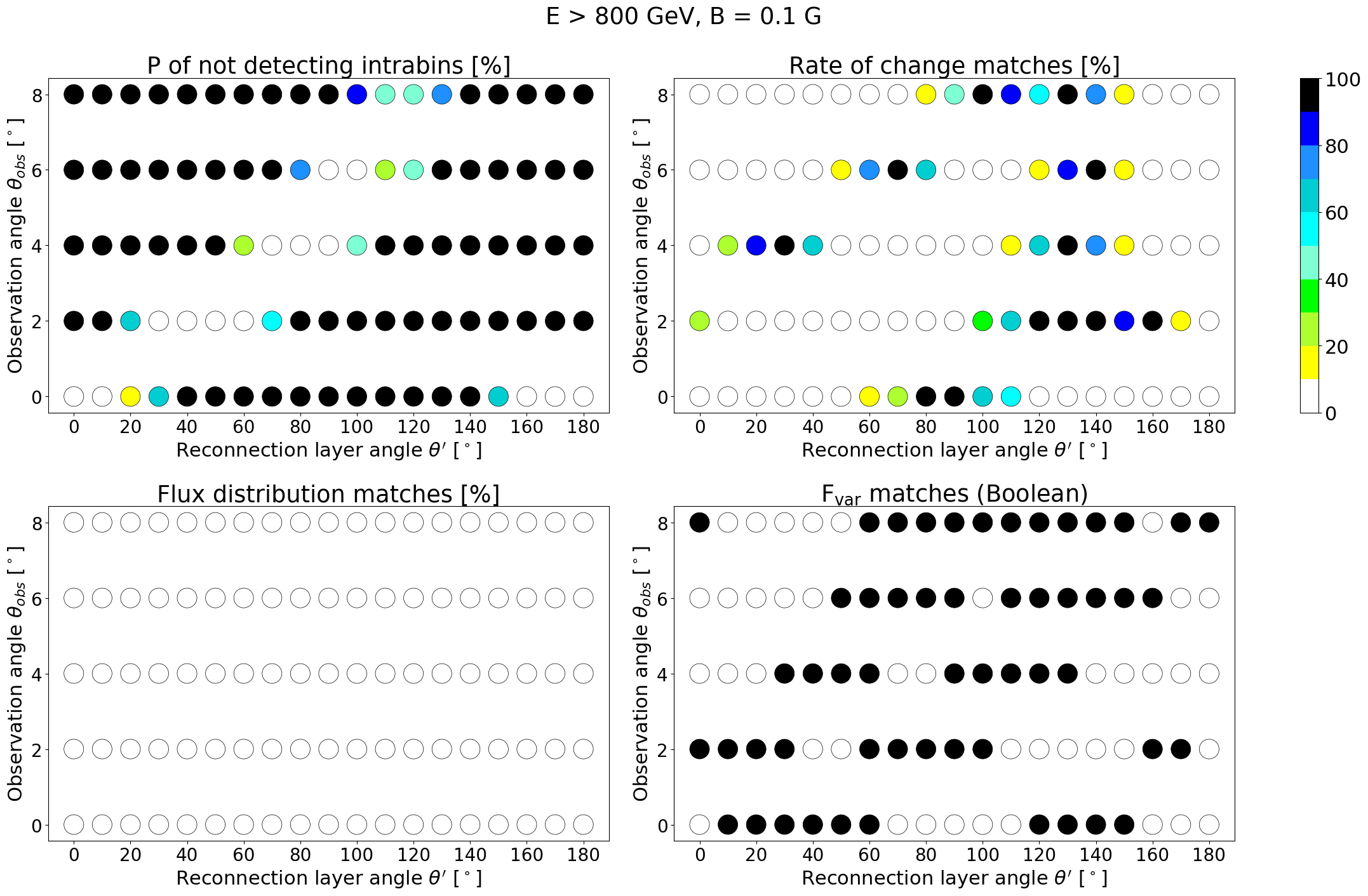}
  \caption{See description in Fig. \ref{fig:ind200}.}
  \label{fig:ind800}
\end{figure*}

The results of the individual tests for the simulations with $B = 0.1$ G described in Sect. \ref{b01sims} are shown here in Figs. \ref{fig:ind400} for 400 - 800 GeV, and \ref{fig:ind800} for $>800$ GeV. The detailed results of all the tests performed are tabulated.

\begin{table*}
\footnotesize
\centering
\caption{Durations and mean fluxes of $B = 0.1$ G simulations.}
\label{tab:B01_results1}
\begin{tabular}{lcccccc} 
\hline\hline
\multicolumn{1}{p{2cm}}{\centering (1) \\ Simulation \\ $(\theta_{obs}, \ \theta ')$} & \multicolumn{3}{p{6cm}}{\centering (2) \\ Tailles duration \\ $[\mathrm{h}]$} & \multicolumn{3}{p{6cm}}{\centering (3) \\ Mean flux \\ $[\mathrm{ph/cm^2/s}]$ \\ $(10^{-9})$} \\ 
\hline
 & 200-400 GeV & 400-800 GeV & $>$800 GeV & 200-400 GeV & 400-800 GeV & $>$800 GeV \\
\hline
(0, 0) & 119 & 118 & 117 & 120 & 60 & 57 \\
(0, 10) & 120 & 119 & 119 & 84 & 42 & 36 \\
(0, 20) & 116 & 115 & 115 & 52 & 26 & 19 \\
(0, 30) & 115 & 115 & 115 & 32 & 16 & 11 \\
(0, 40) & 115 & 116 & 116 & 20 & 9.7 & 6.2 \\
(0, 50) & 183 & 184 & 185 & 8.1 & 3.8 & 1.9 \\
(0, 60) & 178 & 179 & 180 & 6.0 & 2.8 & 1.4 \\
(0, 70) & 175 & 176 & 178 & 4.8 & 2.3 & 1.1 \\
(0, 80) & 167 & 174 & 177 & 4.2 & 1.9 & 0.9 \\
(0, 90) & 176 & 177 & 181 & 4.0 & 1.9 & 0.9 \\
(0, 100) & 170 & 171 & 182 & 4.4 & 2.0 & 0.9 \\
(0, 110) & 172 & 174 & 175 & 5.0 & 2.3 & 1.2 \\
(0, 120) & 172 & 173 & 173 & 6.5 & 3.1 & 1.6 \\
(0, 130) & 173 & 174 & 174 & 8.8 & 4.2 & 2.1 \\
(0, 140) & 174 & 175 & 175 & 13 & 6.6 & 4.1 \\
(0, 150) & 176 & 176 & 177 & 20 & 9.9 & 6.4 \\
(0, 160) & 185 & 185 & 185 & 31 & 16 & 11 \\
(0, 170) & 192 & 192 & 192 & 50 & 24 & 20 \\
(0, 180) & 127 & 126 & 126 & 110 & 52 & 49 \\
\begin{tabular}[c]{@{}c@{}}.\\.\\.\end{tabular} & \begin{tabular}[c]{@{}c@{}}.\\.\\.\end{tabular} & \begin{tabular}[c]{@{}c@{}}.\\.\\.\end{tabular} & \begin{tabular}[c]{@{}c@{}}.\\.\\.\end{tabular} & \begin{tabular}[c]{@{}c@{}}.\\.\\.\end{tabular} & \begin{tabular}[c]{@{}c@{}}.\\.\\.\end{tabular} & \begin{tabular}[c]{@{}c@{}}.\\.\\.\end{tabular} \\
\hline
\end{tabular}
\tablefoot{The observed mean flux values at each energy band are $6.31^{-10} \ \mathrm{ph/cm^2/s}$, 	$2.7810^{-10} \ \mathrm{ph/cm^2/s}$, and $1.5610^{-10} \ \mathrm{ph/cm^2/s}$. The full table is available at the CDS.}
\end{table*}

\begin{table*}
\footnotesize
\centering
\caption{Results of the time scale tests for the simulations with $B = 0.1$ G.}
\label{tab:B01_results2}
\begin{tabular}{lcccccc} 
\hline\hline
\multicolumn{1}{p{2cm}}{\centering (1) \\ Simulation \\ $(\theta_{obs},\theta ')$} & \multicolumn{3}{p{6cm}}{\centering (2) \\ $1-P_{intrabins}$ \\ $[\%]$} & \multicolumn{3}{p{6cm}}{\centering (3) \\ Rate of change matches \\ $[\%]$} \\
\hline
 & 200-400 GeV & 400-800 GeV & $>$800 GeV & 200-400 GeV & 400-800 GeV & $>$800 GeV \\ 
\hline
(0, 0) & 0.0 & 0.2 & 0.0 & 0.0 & 0.0 & 0.0 \\
(0, 10) & 0.4 & 0.0 & 0.0 & 0.0 & 0.0 & 0.0 \\
(0, 20) & 29.2 & 29.2 & 16.3 & 0.0 & 0.0 & 0.0 \\
(0, 30) & 71.4 & 71.4 & 71.4 & 0.0 & 0.0 & 0.0 \\
(0, 40) & 100.0 & 100.0 & 100.0 & 0.0 & 0.0 & 0.0 \\
(0, 50) & 100.0 & 100.0 & 100.0 & 0.3 & 3.8 & 6.5 \\
(0, 60) & 100.0 & 100.0 & 100.0 & 0.8 & 0.4 & 16.3 \\
(0, 70) & 100.0 & 100.0 & 100.0 & 20.6 & 11.9 & 24.9 \\
(0, 80) & 100.0 & 100.0 & 100.0 & 6.8 & 27.7 & 91.2 \\
(0, 90) & 100.0 & 100.0 & 100.0 & 18.5 & 55.1 & 96.8 \\
(0, 100) & 100.0 & 100.0 & 100.0 & 0.0 & 0.5 & 64.5 \\
(0, 110) & 100.0 & 100.0 & 100.0 & 5.7 & 19.5 & 54.8 \\
(0, 120) & 100.0 & 100.0 & 100.0 & 0.0 & 0.4 & 7.0 \\
(0, 130) & 100.0 & 100.0 & 100.0 & 0.0 & 0.0 & 0.8 \\
(0, 140) & 100.0 & 100.0 & 100.0 & 0.0 & 0.0 & 0.0 \\
(0, 150) & 100.0 & 69.0 & 69.2 & 0.0 & 0.0 & 0.0 \\
(0, 160) & 61.1 & 13.2 & 10.8 & 0.0 & 0.0 & 0.0 \\
(0, 170) & 1.6 & 1.2 & 0.0 & 0.0 & 0.0 & 0.0 \\
(0, 180) & 0.0 & 0.0 & 0.0 & 0.0 & 0.0 & 0.0 \\
\begin{tabular}[c]{@{}l@{}}.\\.\\.\end{tabular} & \begin{tabular}[c]{@{}c@{}}.\\.\\.\end{tabular} & \begin{tabular}[c]{@{}c@{}}.\\.\\.\end{tabular} & \begin{tabular}[c]{@{}c@{}}.\\.\\.\end{tabular} & \begin{tabular}[c]{@{}c@{}}.\\.\\.\end{tabular} & \begin{tabular}[c]{@{}c@{}}.\\.\\.\end{tabular} & \begin{tabular}[c]{@{}c@{}}.\\.\\.\end{tabular} \\
\hline
\end{tabular}
\tablefoot{Columns: (1) Simulation name. (2) Probability of not detecting any intrabins in the simulated light curves. (3) Matches of the rate of change per 1000 samples. The full table is available at the CDS.}
\end{table*}

\begin{table*}
\footnotesize
\centering
\caption{Results of the flux amplitude tests for the simulations with $B = 0.1$ G.}
\label{tab:B01_results3}
\begin{tabular}{lcccccc} 
\hline\hline
\multicolumn{1}{p{2cm}}{\centering (1) \\ Simulation \\ $(\theta_{obs},\theta ')$} & \multicolumn{3}{p{6cm}}{\centering (2) \\ Flux
  distribution matches \\ $[\%]$} & \multicolumn{3}{p{6cm}}{\centering (3) \\ $F_{var}$ distribution matches} \\ 
\hline
 & 200-400 GeV & 400-800 GeV & $>$800 GeV & 200-400 GeV & 400-800 GeV & $>$800 GeV \\ 
\hline
(0, 0) & 0.0 & 0.0 & 0.0 & no & no & no \\
(0, 10) & 0.0 & 0.0 & 0.0 & no & no & yes \\
(0, 20) & 0.0 & 0.0 & 0.0 & yes & yes & yes \\
(0, 30) & 0.0 & 0.0 & 0.0 & no & yes & yes \\
(0, 40) & 0.0 & 0.0 & 0.0 & yes & yes & yes \\
(0, 50) & 0.0 & 0.0 & 0.0 & no & yes & yes \\
(0, 60) & 0.0 & 0.0 & 0.0 & yes & yes & yes \\
(0, 70) & 0.0 & 0.0 & 0.0 & yes & yes & no \\
(0, 80) & 0.0 & 0.0 & 0.0 & yes & no & no \\
(0, 90) & 0.0 & 0.0 & 0.0 & yes & yes & no \\
(0, 100) & 0.0 & 0.0 & 0.0 & yes & yes & no \\
(0, 110) & 0.0 & 0.0 & 0.0 & yes & yes & no \\
(0, 120) & 0.0 & 0.0 & 0.0 & no & no & yes \\
(0, 130) & 0.0 & 0.0 & 0.0 & no & no & yes \\
(0, 140) & 0.0 & 0.0 & 0.0 & no & no & yes \\
(0, 150) & 0.0 & 0.0 & 0.0 & no & no & yes \\
(0, 160) & 0.0 & 0.0 & 0.0 & no & no & no \\
(0, 170) & 0.0 & 0.0 & 0.0 & no & no & no \\
(0, 180) & 0.0 & 0.0 & 0.0 & no & no & no \\
\begin{tabular}[c]{@{}c@{}}.\\.\\.\end{tabular} & \begin{tabular}[c]{@{}c@{}}.\\.\\.\end{tabular} & \begin{tabular}[c]{@{}c@{}}.\\.\\.\end{tabular} & \begin{tabular}[c]{@{}c@{}}.\\.\\.\end{tabular} & \begin{tabular}[c]{@{}c@{}}.\\.\\.\end{tabular} & \begin{tabular}[c]{@{}c@{}}.\\.\\.\end{tabular} & \begin{tabular}[c]{@{}c@{}}.\\.\\.\end{tabular} \\
\hline
\end{tabular}
\tablefoot{Columns: (1) Simulation name. (2) Matches of the flux distribution test per a 1000 samples. (3) Matches of the fractional variability test (boolean). The full table is available at the CDS.}
\end{table*}

\begin{table}
\footnotesize
\centering
\caption{Results of the spectral slope distribution test and the spectral slope means of the simulations with $B = 0.1$ G.}
\label{tab:B01_results4}
\begin{tabular}{lccc} 
\hline\hline
\multicolumn{1}{p{2cm}}{\centering (1) \\ Simulation \\ $(\theta_{obs},\theta ')$} & \multicolumn{1}{p{2cm}}{\centering (2) \\ $m$ distribution matches \\ $[\%]$} & \multicolumn{1}{p{1cm}}{\centering (3) \\ $\bar{x}_{m, 1000}$}& \multicolumn{1}{p{1cm}}{\centering (4) \\ $\sigma_{\bar{x}_{m, 1000}}$} \\
\hline
(0, 0) & 0.0 & 0.34 & 0.03 \\
(0, 10) & 0.0 & 0.35 & 0.03 \\
(0, 20) & 0.0 & 0.32 & 0.03 \\
(0, 30) & 0.0 & 0.28 & 0.02 \\
(0, 40) & 0.0 & 0.25 & 0.01 \\
(0, 50) & 0.0 & 0.07 & 0.02 \\
(0, 60) & 0.0 & 0.09 & 0.01 \\
(0, 70) & 0.0 & 0.07 & 0.02 \\
(0, 80) & 0.0 & 0.05 & 0.01 \\
(0, 90) & 0.0 & 0.05 & 0.01 \\
(0, 100) & 0.0 & 0.04 & 0.01 \\
(0, 110) & 0.0 & 0.06 & 0.01 \\
(0, 120) & 0.0 & 0.09 & 0.01 \\
(0, 130) & 0.0 & 0.08 & 0.02 \\
(0, 140) & 0.0 & 0.18 & 0.03 \\
(0, 150) & 0.0 & 0.18 & 0.03 \\
(0, 160) & 0.0 & 0.22 & 0.04 \\
(0, 170) & 0.0 & 0.23 & 0.04 \\
(0, 180) & 0.0 & 0.31 & 0.02 \\
\begin{tabular}[c]{@{}c@{}}.\\.\\.\end{tabular} & \begin{tabular}[c]{@{}c@{}}.\\.\\.\end{tabular} & \begin{tabular}[c]{@{}c@{}}.\\.\\.\end{tabular} &
\begin{tabular}[c]{@{}c@{}}.\\.\\.\end{tabular} \\
\hline
\end{tabular}
\tablefoot{Columns: (1) Simulation name. (2) Matches of the spectral slope distributions. (3) Spectral slope mean of all means of 1000 samples. (4) Standard deviation of $\bar{x}_{m}$. The observed $\bar{x}_{m} = -0.09$ and $\sigma_{\bar{x}_{m}} = 0.28$. The full table is available at the CDS.}
\end{table}

\begin{table*}
\footnotesize
\centering
\caption{Durations and mean fluxes of $B=1$ G simulations.}
\label{tab:B1_results1}
\begin{tabular}{lcccccc} 
\hline\hline
\multicolumn{1}{p{2cm}}{\centering (1) \\ Simulation \\ $(\theta_{obs},\theta ')$} & \multicolumn{3}{p{6cm}}{\centering (2) \\ Tailles duration \\ $[\mathrm{h}]$} & \multicolumn{3}{p{6cm}}{\centering (3) \\ Mean flux \\ $[\mathrm{ph/cm^2/s}]$ \\ $(10^{-10})$} \\ 
\hline
\multicolumn{1}{c}{} & 200-400GeV & 400-800GeV & 800GeV & 200-400Gev & 400-800GeV & 800GeV \\ 
\hline
(0, 0) & 11.4 & 11.4 & 10.8 & 230 & 85 & 28 \\
(0, 10) & 11.7 & 11.6 & 11.6 & 160 & 60 & 17 \\
(0, 20) & 11.3 & 11.4 & 11.3 & 97 & 37 & 8.5 \\
(0, 30) & 11.2 & 11.2 & 11.1 & 57 & 21 & 3.8 \\
(0, 40) & 11.4 & 11.4 & 11.4 & 35 & 13 & 2.1 \\
(0, 50) & 18.0 & 18.1 & 18.1 & 15 & 4.5 & 0.3 \\
(0, 60) & 17.7 & 17.8 & 18.1 & 11 & 3.5 & 0.3 \\
(0, 70) & 17.6 & 17.8 & 18.3 & 8.2 & 2.3 & 0.2 \\
(0, 80) & 17.4 & 17.6 & 17.7 & 7.5 & 2.1 & 0.2 \\
(0, 90) & 17.7 & 17.9 & 18.2 & 6.8 & 1.8 & 0.1 \\
(0, 100) & 17.7 & 17.9 & 18.3 & 7.6 & 2.1 & 0.2 \\
(0, 110) & 17.4 & 17.4 & 17.8 & 8.9 & 2.6 & 0.2 \\
(0, 120) & 17.5 & 17.2 & 17.3 & 12 & 3.5 & 0.3 \\
(0, 130) & 17.1 & 17.2 & 17.2 & 16 & 4.8 & 0.4 \\
(0, 140) & 17.2 & 17.2 & 17.2 & 24 & 8.4 & 1.4 \\
(0, 150) & 17.4 & 17.4 & 17.5 & 38 & 14 & 2.5 \\
(0, 160) & 18.3 & 18.3 & 18.3 & 55 & 20 & 4.3 \\
(0, 170) & 18.9 & 18.9 & 18.8 & 90 & 34 & 9.5 \\
(0, 180) & 12.4 & 12.4 & 8.4 & 210 & 77 & 37 \\
\begin{tabular}[c]{@{}c@{}}.\\.\\.\end{tabular} & \begin{tabular}[c]{@{}c@{}}.\\.\\.\end{tabular} & \begin{tabular}[c]{@{}c@{}}.\\.\\.\end{tabular} & \begin{tabular}[c]{@{}c@{}}.\\.\\.\end{tabular} & \begin{tabular}[c]{@{}c@{}}.\\.\\.\end{tabular} & \begin{tabular}[c]{@{}c@{}}.\\.\\.\end{tabular} & \begin{tabular}[c]{@{}c@{}}.\\.\\.\end{tabular} \\
\hline
\end{tabular}
\tablefoot{The observed mean flux values at each energy band are $6.31^{-10} \ \mathrm{ph/cm^2/s}$, 	$2.7810^{-10} \ \mathrm{ph/cm^2/s}$, and $1.5610^{-10} \ \mathrm{ph/cm^2/s}$. The full table is available at the CDS.}
\end{table*}

\begin{table*}
\footnotesize
\centering
\caption{Results of the time scale tests for the simulations with $B = 1$ G.}
\label{tab:B1_results2}
\begin{tabular}{lcccccc} 
\hline\hline
\multicolumn{1}{p{2cm}}{\centering (1) \\ Simulation \\ $(\theta_{obs},\theta ')$} & \multicolumn{3}{p{6cm}}{\centering (2) \\ $1-P_{intrabins}$ \\ $[\%]$} & \multicolumn{3}{p{6cm}}{\centering (3) \\ Rate of change matches \\ $[\%]$} \\
\hline
 & 200-400 GeV & 400-800 GeV & $>$800 GeV & 200-400 GeV & 400-800 GeV & $>$800 GeV \\ 
\hline
(0, 0) & 0.0 & 0.0 & 0.0 & 0.0 & 0.0 & 0.0 \\
(0, 10) & 0.0 & 0.0 & 0.0 & 0.0 & 0.0 & 0.0 \\
(0, 20) & 0.0 & 0.0 & 0.0 & 0.0 & 0.0 & 0.0 \\
(0, 30) & 0.0 & 0.0 & 0.0 & 0.0 & 0.0 & 0.0 \\
(0, 40) & 0.0 & 0.0 & 0.0 & 0.0 & 0.0 & 0.0 \\
(0, 50) & 100.0 & 0.0 & 40.0 & 0.0 & 0.0 & 90.0 \\
(0, 60) & 100.0 & 100.0 & 100.0 & 0.0 & 0.0 & 100.0 \\
(0, 70) & 100.0 & 100.0 & 100.0 & 10.0 & 50.0 & 0.0 \\
(0, 80) & 100.0 & 100.0 & 100.0 & 40.0 & 90.0 & 30.0 \\
(0, 90) & 80.0 & 80.0 & 100.0 & 20.0 & 100.0 & 0.0 \\
(0, 100) & 80.0 & 80.0 & 100.0 & 0.0 & 60.0 & 20.0 \\
(0, 110) & 90.0 & 90.0 & 100.0 & 0.0 & 60.0 & 10.0 \\
(0, 120) & 100.0 & 70.0 & 100.0 & 0.0 & 0.0 & 70.0 \\
(0, 130) & 90.0 & 0.0 & 90.0 & 0.0 & 0.0 & 100.0 \\
(0, 140) & 0.0 & 0.0 & 0.0 & 0.0 & 0.0 & 30.0 \\
(0, 150) & 0.0 & 0.0 & 0.0 & 0.0 & 0.0 & 0.0 \\
(0, 160) & 0.0 & 0.0 & 0.0 & 0.0 & 0.0 & 3.3 \\
(0, 170) & 0.0 & 0.0 & 0.0 & 0.0 & 0.0 & 0.0 \\
(0, 180) & 0.0 & 0.0 & 0.0 & 0.0 & 0.0 & 0.0 \\
\begin{tabular}[c]{@{}c@{}}.\\.\\.\end{tabular} & \begin{tabular}[c]{@{}c@{}}.\\.\\.\end{tabular} & \begin{tabular}[c]{@{}c@{}}.\\.\\.\end{tabular} & \begin{tabular}[c]{@{}c@{}}.\\.\\.\end{tabular} & \begin{tabular}[c]{@{}c@{}}.\\.\\.\end{tabular} & \begin{tabular}[c]{@{}c@{}}.\\.\\.\end{tabular} & \begin{tabular}[c]{@{}c@{}}.\\.\\.\end{tabular} \\
\hline
\end{tabular}
\tablefoot{The full table is available at the CDS.}
\end{table*}

\begin{table*}
\footnotesize
\centering
\caption{Results of the flux amplitude tests for the simulations with $B = 1$ G.}
\label{tab:B1_results3}
\begin{tabular}{lcccccc} 
\hline\hline
\multicolumn{1}{p{2cm}}{\centering (1) \\ Simulation \\ $(\theta_{obs},\theta ')$} & \multicolumn{3}{p{6cm}}{\centering (2) \\ Flux
  distribution matches \\ $[\%]$} & \multicolumn{3}{p{6cm}}{\centering (3) \\ $F_{var}$ distribution matches \\ $[\%]$} \\ 
\hline
 & 200-400 GeV & 400-800 GeV & $>$800 GeV & 200-400 GeV & 400-800 GeV & $>$800 GeV \\ 
\hline
(0, 0) & 0.0 & 0.0 & 0.0 & 0.0 & 0.0 & 0.0 \\
(0, 10) & 0.0 & 0.0 & 0.0 & 0.0 & 0.0 & 0.0 \\
(0, 20) & 0.0 & 0.0 & 0.0 & 0.0 & 0.0 & 0.0 \\
(0, 30) & 0.0 & 0.0 & 0.0 & 0.0 & 0.0 & 0.0 \\
(0, 40) & 0.0 & 0.0 & 0.0 & 0.0 & 0.0 & 0.0 \\
(0, 50) & 0.0 & 0.0 & 8.3 & 0.0 & 0.0 & 0.0 \\
(0, 60) & 0.0 & 6.7 & 6.7 & 0.0 & 0.0 & 0.0 \\
(0, 70) & 6.7 & 0.0 & 0.0 & 0.0 & 3.3 & 23.3 \\
(0, 80) & 0.0 & 6.7 & 0.0 & 0.0 & 0.0 & 0.0 \\
(0, 90) & 8.3 & 8.3 & 0.0 & 0.0 & 0.0 & 0.0 \\
(0, 100) & 3.3 & 6.7 & 0.0 & 0.0 & 0.0 & 0.0 \\
(0, 110) & 0.0 & 1.7 & 0.0 & 0.0 & 0.0 & 0.0 \\
(0, 120) & 0.0 & 0.0 & 0.0 & 0.0 & 0.0 & 0.0 \\
(0, 130) & 0.0 & 0.0 & 0.0 & 0.0 & 0.0 & 0.0 \\
(0, 140) & 0.0 & 0.0 & 0.0 & 0.0 & 0.0 & 0.0 \\
(0, 150) & 0.0 & 0.0 & 0.0 & 0.0 & 0.0 & 0.0 \\
(0, 160) & 0.0 & 0.0 & 0.0 & 0.0 & 0.0 & 0.0 \\
(0, 170) & 0.0 & 0.0 & 0.0 & 0.0 & 0.0 & 0.0 \\
(0, 180) & 0.0 & 0.0 & 0.0 & 0.0 & 0.0 & 0.0 \\
\begin{tabular}[c]{@{}c@{}}.\\.\\.\end{tabular} & \begin{tabular}[c]{@{}c@{}}.\\.\\.\end{tabular} & \begin{tabular}[c]{@{}c@{}}.\\.\\.\end{tabular} & \begin{tabular}[c]{@{}c@{}}.\\.\\.\end{tabular} & \begin{tabular}[c]{@{}c@{}}.\\.\\.\end{tabular} & \begin{tabular}[c]{@{}c@{}}.\\.\\.\end{tabular} & \begin{tabular}[c]{@{}c@{}}.\\.\\.\end{tabular} \\
\hline
\end{tabular}
\tablefoot{The full table is available at the CDS.}
\end{table*}

\begin{table*}
\footnotesize
\centering
\caption{Results of the spectral slope distribution test and the spectral slope means of the simulations with $B = 1$ G.}
\label{tab:B1_results4}
\begin{tabular}{lccc} 
\hline\hline
\multicolumn{1}{p{2cm}}{\centering (1) \\ Simulation \\ $(\theta_{obs},\theta ')$} & \multicolumn{1}{p{2cm}}{\centering (2) \\ $m$ distribution matches \\ $[\%]$} & \multicolumn{1}{p{1cm}}{\centering (3) \\ $\bar{x}_{m}$}& \multicolumn{1}{p{1cm}}{\centering (4) \\ $\sigma_{\bar{x}_{m}}$} \\
\hline
(0, 0) & 16.7 & -0.39 & 0.19 \\
(0, 10) & 0.0 & -0.45 & 0.21 \\
(0, 20) & 0.0 & -0.57 & 0.26 \\
(0, 30) & 0.0 & -0.69 & 0.12 \\
(0, 40) & 0.0 & -0.72 & 0.05 \\
(0, 50) & 0.0 & -1.32 & 0.03 \\
(0, 60) & 0.0 & -1.27 & 0.02 \\
(0, 70) & 0.0 & -1.34 & 0.03 \\
(0, 80) & 0.0 & -1.36 & 0.02 \\
(0, 90) & 0.0 & -1.40 & 0.02 \\
(0, 100) & 0.0 & -1.38 & 0.02 \\
(0, 110) & 0.0 & -1.31 & 0.02 \\
(0, 120) & 0.0 & -1.27 & 0.01 \\
(0, 130) & 0.0 & -1.30 & 0.02 \\
(0, 140) & 0.0 & -0.88 & 0.08 \\
(0, 150) & 0.0 & -0.81 & 0.10 \\
(0, 160) & 0.0 & -0.78 & 0.11 \\
(0, 170) & 0.0 & -0.64 & 0.13 \\
(0, 180) & 16.7 & -0.36 & 0.28 \\
\begin{tabular}[c]{@{}c@{}}.\\.\\.\end{tabular} & \begin{tabular}[c]{@{}c@{}}.\\.\\.\end{tabular} & \begin{tabular}[c]{@{}c@{}}.\\.\\.\end{tabular} &  \begin{tabular}[c]{@{}c@{}}.\\.\\.\end{tabular} \\
\hline
\end{tabular}
\tablefoot{The observed $\bar{x}_{m} = -0.09$ and $\sigma_{\bar{x}_{m}} = 0.28$. The full table is available at the CDS.}
\end{table*}

\begin{table*}
\footnotesize
\centering
\caption{Durations and mean fluxes of B=10G simulations.}
\label{tab:B10_results1}
\begin{tabular}{lcccccc} 
\hline\hline
\multicolumn{1}{p{2cm}}{\centering (1) \\ Simulation \\ $(\theta_{obs},\theta ')$} & \multicolumn{3}{p{6cm}}{\centering (2) \\ Tailles duration \\ $[\mathrm{h}]$} & \multicolumn{3}{p{6cm}}{\centering (3) \\ Mean flux \\ $[\mathrm{ph/cm^2/s}]$ \\ $(10^{-11})$} \\ 
\hline
 & 200-400GeV & 400-800GeV & 800GeV & 200-400GeV & 400-800GeV & 800GeV \\ 
\hline
(0, 0) & 0.88 & 0.86 & 0.74 & 110 & 18 & 0.9 \\
(0, 10) & 0.88 & 0.87 & 0.67 & 75 & 11 & 0.4 \\
(0, 20) & 0.87 & 0.87 & 0.86 & 46 & 5.2 & 0.01 \\
(0, 30) & 0.86 & 0.86 & - & 27 & 2.5 & - \\
(0, 40) & 0.87 & 0.86 & - & 20 & 2.0 & - \\
(0, 50) & 1.54 & 1.61 & - & 4.3 & 0.06 & - \\
(0, 60) & 1.52 & 1.54 & - & 3.7 & 0.1 & - \\
(0, 70) & 1.49 & 1.91 & - & 2.6 & 0.04 & - \\
(0, 80) & 1.50 & 1.91 & - & 2.0 & 0.01 & - \\
(0, 90) & 1.54 & 1.73 & - & 1.9 & 0.01 & - \\
(0, 100) & 1.53 & 1.90 & - & 2.1 & 0.01 & - \\
(0, 110) & 1.47 & 1.69 & - & 2.7 & 0.06 & - \\
(0, 120) & 1.44 & 1.45 & - & 4.3 & 0.2 & - \\
(0, 130) & 1.46 & 1.70 & - & 5.1 & 0.1 & - \\
(0, 140) & 1.46 & 1.45 & - & 11 & 1.0 & - \\
(0, 150) & 1.48 & 1.48 & - & 16 & 1.4 & - \\
(0, 160) & 1.55 & 1.63 & 0.71 & 26 & 2.8 & 0.04 \\
(0, 170) & 1.63 & 1.63 & 0.93 & 40 & 5.7 & 0.3 \\
(0, 180) & 1.64 & 0.93 & 0.43 & 59 & 14 & 1.2 \\
\begin{tabular}[c]{@{}l@{}}.\\.\\.\end{tabular} & \begin{tabular}[c]{@{}c@{}}.\\.\\.\end{tabular} & \begin{tabular}[c]{@{}c@{}}.\\.\\.\end{tabular} & \begin{tabular}[c]{@{}c@{}}.\\.\\.\end{tabular} & \begin{tabular}[c]{@{}c@{}}.\\.\\.\end{tabular} & \begin{tabular}[c]{@{}c@{}}.\\.\\.\end{tabular} & \begin{tabular}[c]{@{}c@{}}.\\.\\.\end{tabular} \\
\hline
\end{tabular}
\tablefoot{The full table is available at the CDS.}
\end{table*}

\begin{table}
\footnotesize
\centering
\caption{Results of the intrabin variability test for the simulations with $B=10$ G.}
\label{tab:B10_results2}
\begin{tabular}{lccc} 
\hline\hline
\multicolumn{1}{p{2cm}}{\centering (1) \\ Simulation \\ $(\theta_{obs},\theta ')$} & \multicolumn{3}{p{6cm}}{\centering (2) \\ $1-P_{intrabins}$ \\ $[\%]$} \\
\hline
 & 200-400GeV & 400-800GeV & 800GeV \\ 
\hline
(0, 0) & 100.0 & 100.0 & 100.0 \\
(0, 10) & 100.0 & 100.0 & 100.0 \\
(0, 20) & 100.0 & 100.0 & 100.0 \\
(0, 30) & 100.0 & 100.0 & - \\
(0, 40) & 100.0 & 100.0 & - \\
(0, 50) & 100.0 & 100.0 & - \\
(0, 60) & 100.0 & 100.0 & - \\
(0, 70) & 100.0 & 100.0 & - \\
(0, 80) & 100.0 & 100.0 & - \\
(0, 90) & 100.0 & 100.0 & - \\
(0, 100) & 100.0 & 100.0 & - \\
(0, 110) & 100.0 & 100.0 & - \\
(0, 120) & 100.0 & 100.0 & - \\
(0, 130) & 100.0 & 100.0 & - \\
(0, 140) & 100.0 & 100.0 & - \\
(0, 150) & 100.0 & 100.0 & - \\
(0, 160) & 100.0 & 100.0 & 100.0 \\
(0, 170) & 100.0 & 100.0 & 100.0 \\
(0, 180) & 100.0 & 100.0 & 100.0 \\
\begin{tabular}[c]{@{}l@{}}.\\.\\.\end{tabular} & \begin{tabular}[c]{@{}c@{}}.\\.\\.\end{tabular} & \begin{tabular}[c]{@{}c@{}}.\\.\\.\end{tabular} & \begin{tabular}[c]{@{}c@{}}.\\.\\.\end{tabular} \\
\hline
\end{tabular}
\tablefoot{The full table is available at the CDS.}
\end{table}

\end{appendix}

\end{document}